\documentclass[longauth]{aa} % for the long lists of affiliations 
\usepackage{graphicx,natbib}
\usepackage{txfonts}
\usepackage{longtable}
\begin{document}
   \title{Variability of the blazar \object{4C 38.41} (\object{B3 1633+382}) from GHz frequencies to GeV energies\thanks{The radio-to-optical 
   data collected by the GASP-WEBT collaboration are stored in the GASP-WEBT archive; 
   for questions regarding their availability,
   please contact the WEBT President Massimo Villata ({\tt villata@oato.inaf.it}).}
}

   \subtitle{} 

   \author{C.~M.~Raiteri              \inst{ 1}
   \and   M.~Villata                  \inst{ 1}
   \and   P.~S.~Smith                 \inst{ 2}
   \and   V.~M.~Larionov              \inst{ 3,4,5}
   \and   J.~A.~Acosta-Pulido         \inst{ 6,7}
   \and   M.~F.~Aller                 \inst{ 8}
   \and   F.~D'Ammando                \inst{ 9}
   \and   M.~A.~Gurwell               \inst{10}
   \and   S.~G.~Jorstad               \inst{11,3}
   \and   M.~Joshi                    \inst{11}
   \and   O.~M.~Kurtanidze            \inst{12,13,14,15}
   \and   A.~L\"ahteenm\"aki          \inst{16}
   \and   D.~O.~Mirzaqulov            \inst{17}
   \and   I.~Agudo                    \inst{18,11}
   \and   H.~D.~Aller                 \inst{ 8}
   \and   M.~J.~Ar\'evalo             \inst{ 6,7}
   \and   A.~A.~Arkharov              \inst{ 4}
   \and   U.~Bach                     \inst{19}
   \and   E.~Ben\'itez                \inst{20}
   \and   A.~Berdyugin                \inst{21}
   \and   D.~A.~Blinov                \inst{ 3}
   \and   K.~Blumenthal               \inst{11}
   \and   C.~S.~Buemi                 \inst{22}
   \and   A.~Bueno                    \inst{ 6,7}
   \and   T.~M.~Carleton              \inst{ 2}
   \and   M.~I.~Carnerero             \inst{ 6,7,1}
   \and   D.~Carosati                 \inst{23,24}
   \and   C.~Casadio                  \inst{18}
   \and   W.~P.~Chen                  \inst{25}
   \and   A.~Di Paola                 \inst{26}
   \and   M.~Dolci                    \inst{27}
   \and   N.~V.~Efimova               \inst{ 3,4}
   \and   Sh.~A.~Ehgamberdiev         \inst{17}
   \and   J.~L.~G\'omez               \inst{18}
   \and   A.~I.~Gonz\'alez            \inst{ 7}
   \and   V.~A.~Hagen-Thorn           \inst{ 3,5}
   \and   J.~Heidt                    \inst{14}
   \and   D.~Hiriart                  \inst{28}
   \and   Sh.~Holikov                 \inst{17}
   \and   T.~S.~Konstantinova         \inst{ 3}
   \and   E.~N.~Kopatskaya            \inst{ 3}
   \and   E.~Koptelova                \inst{25,29}
   \and   S.~O.~Kurtanidze            \inst{12}
   \and   E.~G.~Larionova             \inst{ 3}
   \and   L.~V.~Larionova             \inst{ 3}
   \and   J.~Le\'on-Tavares           \inst{16}
   \and   P.~Leto                     \inst{22}
   \and   H.~C.~Lin                   \inst{25}
   \and   E.~Lindfors                 \inst{21}
   \and   A.~P.~Marscher              \inst{11}
   \and   I.~M.~McHardy               \inst{30}
   \and   S.~N.~Molina                \inst{18}
   \and   D.~A.~Morozova              \inst{ 3}
   \and   R.~Mujica                   \inst{31}
   \and   M.~G.~Nikolashvili          \inst{12}
   \and   K.~Nilsson                  \inst{32}
   \and   E.~P.~Ovcharov              \inst{33}
   \and   N.~Panwar                   \inst{25}
   \and   M.~Pasanen                  \inst{21}
   \and   I.~Puerto-Gimenez           \inst{ 6,34}
   \and   R.~Reinthal                 \inst{21}
   \and   G.~M.~Richter               \inst{13}
   \and   J.~A.~Ros                   \inst{35}
   \and   T.~Sakamoto                 \inst{36,37}
   \and   R.~D.~Schwartz              \inst{38}
   \and   A.~Sillanp\"a\"a            \inst{21}
   \and   N.~Smith                    \inst{39}
   \and   L.~O.~Takalo                \inst{21}
   \and   J.~Tammi                    \inst{16}
   \and   B.~Taylor                   \inst{11,40}
   \and   C.~Thum                     \inst{41}
   \and   M.~Tornikoski               \inst{16}
   \and   C.~Trigilio                 \inst{22}
   \and   I.~S.~Troitsky              \inst{ 3}
   \and   G.~Umana                    \inst{22}
   \and   A.~T.~Valcheva              \inst{33}
   \and   A.~E.~Wehrle                \inst{42}
 }
  
   \institute{
 % 1
          INAF, Osservatorio Astronomico di Torino, Italy                                                     
 %           2
   \and   Steward Observatory, University of Arizona, Tucson, AZ, USA                                         
 %           3
   \and   Astron.\ Inst., St.-Petersburg State Univ., Russia                                                  
 %           4
   \and   Pulkovo Observatory, St.-Petersburg, Russia                                                         
 %           5
   \and   Isaac Newton Institute of Chile, St.-Petersburg Branch                                              
 %           6
   \and   Instituto de Astrofisica de Canarias (IAC), La Laguna, Tenerife, Spain                              
 %           7
   \and   Departamento de Astrofisica, Universidad de La Laguna, La Laguna, Tenerife, Spain                   
 %           8
   \and   Department of Astronomy, University of Michigan, MI, USA                                            
 %           9
   \and   Dip.\ di Fisica, Universit\`a degli Studi di Perugia, Perugia, Italy                                
 %          10
   \and   Harvard-Smithsonian Center for Astrophysics, Cambridge, MA, USA                                     
 %          11
   \and   Institute for Astrophysical Research, Boston University, MA, USA                                    
 %          12
   \and   Abastumani Observatory, Mt. Kanobili, Abastumani, Georgia                                           
 %          13
   \and   Astrophysikalisches Institut Potsdam, An der Sternwarte 16, Potsdam, Germany                        
 %          14
   \and   Landessternwarte Heidelberg-K\"onigstuhl, Heidelberg, Germany                                       
 %          15
   \and   Engelhardt Astronomical Observatory, Kazan Federal Univ., Tatarstan, Russia                         
 %          16
   \and   Aalto University Mets\"ahovi Radio Observatory, Kylm\"al\"a, Finland                                
 %          17
   \and   Maidanak Observatory of the Ulugh Beg Astronomical Institute, Uzbekistan                            
 %          18
   \and   Instituto de Astrof\'{i}sica de Andaluc\'{i}a, CSIC, Granada, Spain                                 
 %          19
   \and   Max-Planck-Institut f\"ur Radioastronomie, Bonn, Germany                                            
 %          20
   \and   Instituto de Astronom\'ia, Universidad Nacional Aut\'onoma de M\'exico, M\'exico DF, M\'exico       
 %          21
   \and   Tuorla Observatory, Dept.\ of Physics and Astronomy, Univ.\ of Turku, Piikki\"o, Finland            
 %          22
   \and   INAF, Osservatorio Astrofisico di Catania, Italy                                                    
 %          23
   \and   EPT Observatories, Tijarafe, La Palma, Spain,                                                       
 %          24
   \and   INAF, TNG Fundaci\'on Galileo Galilei, La Palma, Spain                                              
 %          25
   \and   Graduate Inst.\ of Astronomy, National Central Univ., Jhongli, Taiwan                               
 %          26
   \and   INAF, Osservatorio Astronomico di Roma, Italy                                                       
 %          27
   \and   INAF, Osservatorio Astronomico di Collurania Teramo, Italy                                          
 %          28
   \and   Instituto de Astronomía, Universidad Nacional Aut\'onoma de M\'exico, Ensenada, M\'exico           
 %          29
   \and   Department of Physics, National Taiwan University, Taipei, Taiwan                                   
 %          30
   \and   Dept.\ of Physics and Astronomy, Univ.\ of Southampton, Southampton, United Kingdom                 
 %          31
   \and   Instituto Nacional de Astrofisica, \'Optica y Electr\'onica, Puebla, M\'exico                       
 %          32
   \and   Finnish Centre for Astronomy with ESO (FINCA), University of Turku, Piikki\"o, Finland              
 %          33
   \and   Department of Astronomy, University of Sofia, Sofia, Bulgaria                                       
 %          34
   \and   European Space Astronomy Centre (INSA-ESAC), Villanueva de la Cañada, Madrid, Spain                
 %          35
   \and   Agrupaci\'o Astron\`omica de Sabadell, Spain                                                        
 %          36
   \and   CRESST and Astroparticle Physics Laboratory NASA/GSFC, Greenbelt, MD, USA                           
 %          37
   \and   Dept.\ of Physics, Univ.\ of Maryland, Baltimore County, Baltimore, MD, USA                         
 %          38
   \and   Galaxy View Observatory, Sequim, Washington, USA                                                    
 %          39
   \and   Cork Institute of Technology, Cork, Ireland                                                         
 %          40
   \and   Lowell Observatory, Flagstaff, AZ, USA                                                              
 %          41
   \and   Institut de Radio Astronomie Millim\'{e}trique, St.\ Martin d'H\`{e}res, France                     
 %          42
   \and   Space Science Institute, Boulder, CO, USA                                                           
 }

 \date{}

  \abstract
  % context heading (optional)
  % {} leave it empty if necessary  
   {After years of modest optical activity, the quasar-type blazar \object{4C 38.41} (\object{B3 1633+382}) experienced a large outburst in 2011, which was detected throughout the entire electromagnetic spectrum, renewing interest in this source.}
  % aims heading (mandatory)
   {We present the results of low-energy multifrequency monitoring by the GASP project of the WEBT consortium and collaborators, as well as those of spectropolarimetric/spectrophotometric monitoring at the Steward Observatory. We also analyse high-energy observations of the {{\it Swift}} and {{\it Fermi}} satellites. This combined study aims to provide insights into the source broad-band emission and variability properties.}
  % methods heading (mandatory)
   {We assemble optical, near-infrared, millimetre, and radio light curves and investigate their features and correlations. In the optical, we also analyse the spectroscopic and polarimetric properties of the source. We then compare the low-energy emission behaviour with that at high energies.}
  % results heading (mandatory)
   {In the optical--UV band, several results indicate that there is a contribution from a quasi-stellar-object (QSO) like emission component, in addition to both variable and polarised jet emission. In the optical, the source is redder-when-brighter, at least for $R \ga 16$. The optical spectra display broad emission lines, whose flux is constant in time. The observed degree of polarisation increases with flux and is higher in the red than the blue. The spectral energy distribution reveals a bump peaking around the $U$ band. The unpolarised emission component is likely thermal radiation from the accretion disc that dilutes the jet polarisation. We estimate its brightness to be $R_{\rm QSO} \sim 17.85$--18 and derive the intrinsic jet polarisation degree. 
We find no clear correlation between the optical and radio light curves, while the correlation between the optical and $\gamma$-ray flux apparently fades in time, likely because of an increasing optical to $\gamma$-ray flux ratio.}
  % conclusions heading (optional), leave it empty if necessary 
   {As suggested for other blazars, the long-term variability of 4C 38.41 can be interpreted in terms of an inhomogeneous bent jet, where different emitting regions can change their alignment with respect to the line of sight, leading to variations in the Doppler factor $\delta$. 
Under the hypothesis that in the period 2008--2011 all the $\gamma$-ray and optical variability on a one-week timescale were due to changes in $\delta$, this would range between $\sim 7$ and $\sim 21$. If the variability were caused by changes in the viewing angle $\theta$ only, then $\theta$ would go from $\sim 2.6 \degr$ to $\sim 5 \degr$. Variations in the viewing angle would also account for the dependence of the polarisation degree on the source brightness in the framework of a shock-in-jet model.
}


   \maketitle
%
%________________________________________________________________

\section{Introduction}

The Compton Gamma Ray Observatory (CGRO) launched in 1991 revealed 271 $\gamma$-ray sources, one fourth of which were identified as blazars \citep{har99}, i.e.\ active galactic nuclei (AGNs) showing the most extreme properties.
They are indeed characterised by violent activity at almost all frequencies on different timescales, from long-term flux changes to intraday variability (IDV), by high radio to optical polarisation, and by the superluminal motions of radio knots. The observations can be explained by assuming that their emission is relativistically beamed, which occurs if the emitting plasma jet, produced by a supermassive black hole fed by an accretion disc, is directed toward us. 
The polarised, low-energy emission (from the radio to the optical--X-ray band) is likely synchrotron radiation produced by relativistic electrons in the jet, while the high-energy emission (from the X-ray to the $\gamma$-ray frequencies) is usually interpreted as the result of inverse-Compton scattering of low-energy photons off the same relativistic electrons. 
Whether these low-energy photons come from the jet itself (synchrotron self-Compton, or SSC, models) or from the AGN environment (external Compton, or EC, models), and in the latter case whether they come from the accretion disc, the broad line region, or the dusty torus, is still a matter of debate.

Blazars include flat spectrum radio quasars (FSRQs) and BL Lacertae objects. 
Frequently FSRQs exhibit quasi-stellar-object (QSO) like broad emission lines, as well as a blue and unpolarized continuum that is presumed to be the signature of the thermal emission from the accretion disc, the so-called ``big blue bump" \citep[e.g.][]{wil92,pia99,rai07b,dam09}. In contrast, BL Lacertae objects may have by definition at most weak lines, even if sometimes these sources challenge their classification \citep[e.g.][]{ver95}.

Researchers are trying to understand the structure of blazars and the mechanisms behind their emission and variability by analysing the multifrequency behaviour of these objects extended over the broadest possible energy range with data sampling and quality that is as high as possible \citep[see e.g.][]{mar10,jor10,agu11a,agu11b}.
The international collaboration known as the Whole Earth Blazar Telescope (WEBT)\footnote{\tt http://www.oato.inaf.it/blazars/webt/} was created in 1997 for this purpose, and involves tens of optical, radio, and near-infrared observatories. The WEBT campaigns have produced low-frequency light curves of extraordinary sampling, and these results have often been analysed in conjunction with high-frequency data from satellites \citep[see e.g.][and references therein]{vil06,rai08a,rai08c,lar08,boe09,vil09a,rai09}.

Renewed interest in blazars occurred with the launch of the new-generation $\gamma$-ray satellites Astrorivelatore Gamma a Immagini Leggero (AGILE) in 2007, and particularly with that of {\it Fermi} (formerly GLAST) in 2008, which operates in survey mode. To acquire low-energy data to compare with the high-energy observations of AGILE and {\it Fermi}, in 2007 the WEBT started the GLAST-AGILE Support Program \citep[GASP; see e.g.][]{vil08,vil09b,dam09,rai10,rai11,dam11}.
Since 2008, a ground-based monitoring programme has been running at the Steward Observatory \citep{smi09}, providing support for the {\it Fermi} gamma-ray telescope.
It uses the 2.3 m Bok and 1.54 m Kuiper telescopes with the SPOL spectropolarimeter \citep{sch92} and provides publicly available spectropolarimetry, spectrophotometry, and calibrated broad-band flux measurements for about 40 blazars\footnote{\tt http://james.as.arizona.edu/$\sim$psmith/Fermi/}.

In this paper, we study the multifrequency behaviour of one blazar, namely the optically violent variable (OVV) FSRQ 4C 38.41 (1633+382) at redshift $z=1.814$, which belongs to the target list of both the GASP and the Steward Observatory monitoring programmes. 
The period of observations presented here includes a big optical outburst in 2011 \citep{rai11_atel} that was also detected at $\gamma$-ray energies by {\it Fermi} \citep{szo11}, and in the X-ray and UV bands by the {\it Swift} satellite, providing an excellent opportunity to study the source variability over most of the electromagnetic spectrum.

4C 38.41 had already been observed in $\gamma$-rays by the Energetic Gamma Ray Experiment Telescope (EGRET) instrument onboard CGRO several times, and from these data its emission was found to vary significantly on a day timescale \citep{mat93}. 
The maximum flux detected by EGRET above 100 MeV was $(107.5 \pm 9.6) \times 10^{-8} \rm \, ph \, cm^{-2} \, s^{-1}$ in mid September 1991\footnote{\tt http://cossc.gsfc.nasa.gov/cossc/egret/}.
AGILE observed 4C 38.41 several times. In particular, a preliminary analysis of the AGILE Gamma Ray Imaging Detector (GRID) data between 2009 December 1 and 2010 November 30 yielded a $\gamma$-ray flux $F_{\rm E>100\,MeV} = (28 \pm 5) \times 10^{-8} \rm \, ph \, cm^{-2} \, s^{-1}$.
This flux is consistent with the value reported in the Second {\it Fermi} Large Area Telescope (LAT) Catalog \citep{nol12}.
A refined analysis of the AGILE/GRID data, along with a search for possible short-term variability, will be presented in Vercellone et al.\ (2012, in preparation).
Multifrequency observations of 4C 38.41 during the $\gamma$-ray flares observed by {\it Fermi} in 2009--2010, including Very Long Baseline Array (VLBA) images, was presented by \citet{jor11}. They conclude that high states at $\gamma$-ray energies are due to interaction between a disturbance travelling down the jet and the 43 GHz VLBI core.

   \begin{figure*}
   \centering
   \includegraphics{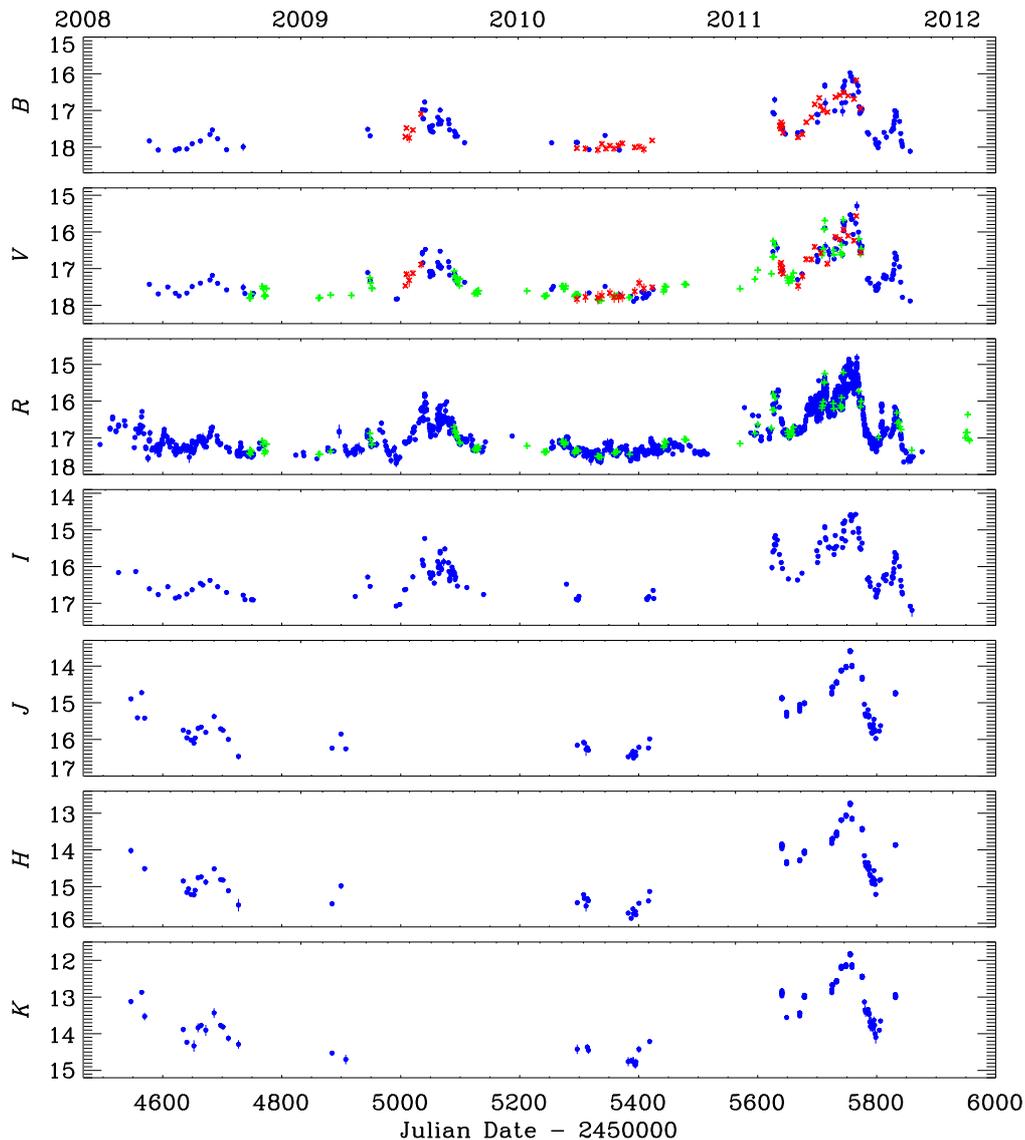}
   \caption{Optical and near-infrared light curves of 4C 38.41 in 2008--2011 built with GASP-WEBT data (blue dots), {\it Swift}-UVOT data (red crosses), and data from the Steward Observatory (green plus signs). 
The UVOT $v$-band points have been shifted by $-0.1$ mag to match the ground-based data.
}
    \label{ottico}
    \end{figure*}

\section{Optical and near-infrared photometry}
\label{sec_opt}

As mentioned in the introduction, the GASP was started in 2007 to perform long-term optical-to-radio monitoring of selected $\gamma$-loud blazars during the observations of the AGILE and {\it Fermi} $\gamma$-ray satellites. In the optical, GASP data are collected in the $R$ band only. However, for this paper we also added data taken by the GASP-WEBT observers in the $B$, $V$, and $I$ bands. 
Calibration of 4C 38.41 was achieved by performing differential photometry with respect to stars A and B in the source field, whose standard magnitudes in the $B$, $V$, and $R$ bands were derived by \citet{vil97}. We obtained standard magnitudes in the $I$ band from the Sloan Digital Sky Survey's $r$ and $i$ photometry, after applying the transformations of \citet{cho08}. In this way, we derived $I=15.20 \pm 0.05$ for star A and $I=15.06 \pm 0.06$ for star B.
The GASP near-infrared (near-IR) data are collected in the $J$, $H$, and $K$ bands, where calibration is performed according to the photometric sequences obtained with the AZT-24 telescope at Campo Imperatore\footnote{\tt http://www.astro.spbu.ru/staff/vlar/NIRlist.html}.

The optical and near-IR light curves of 4C 38.41 in the period 2008--2011 are shown in Fig.\ \ref{ottico}. The GASP-WEBT observations were performed by the following observatories:
Abastumani, 
Calar Alto\footnote{Calar Alto data were acquired as part of the MAPCAT project {\tt http://www.iaa.es/~iagudo/research/MAPCAT}}, 
Campo Imperatore, 
Crimean,
Galaxy View, 
Goddard (GRT),
Lowell (Perkins),
Lulin, 
Mount (Mt.) Maidanak, 
Roque (KVA and Liverpool),
Rozhen,
Sabadell,
San Pedro Martir,
St. Petersburg,
Teide (IAC80 and TCS),
Tijarafe, and
Torino.
The $V$ and $R$-band light curves are complemented by data taken at the Steward Observatory in the framework of the monitoring programme in support of the {\it Fermi} observations (see Sect.\ \ref{sec_stw}).
The figure also displays optical data acquired by the UVOT instrument onboard the {\it Swift} satellite (see Sect.\ \ref{sec_uvot}).

Light curves were carefully checked and cleaned by reducing the data scattering through binning data close in time from the same telescope, when possible, and by removing clear outliers as well as data with errors greater than 0.2 mag.

Substantial variability characterises the source behaviour, especially in 2009 and 2011.
The near-IR light curves were more coarsely sampled than the optical ones, but show the same trend.
In particular, they confirm the sharp brightness fall following the peak of the 2011 outburst.
The range of magnitude spanned over the whole period is 2.1--3.0 mag, depending on the band,
and in general appears to be larger at lower frequencies. This is a well-known feature of FSRQs, which is ascribed to a mixture of variable emission from a jet and almost constant emission from an accretion disc.

\subsection{Colour analysis}
\begin{figure}
   \centering
   \resizebox{\hsize}{!}{\includegraphics{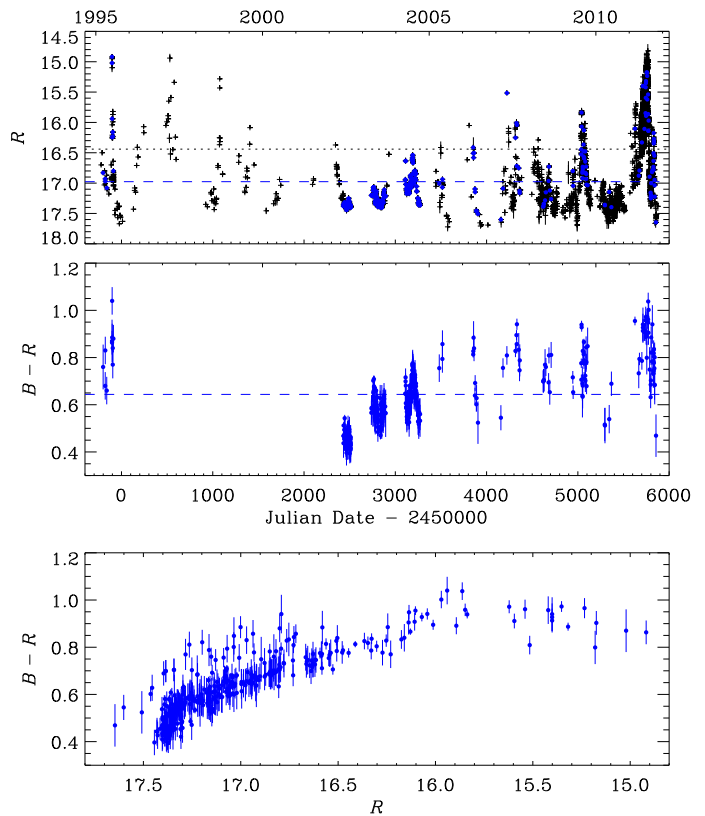}}
   \caption{The $R$-band light curve of 4C 38.41 in 1995--2011 (top panel). Blue dots indicate the data used to derive the $B-R$ colour indices shown both as a function of time (middle panel) and as a function of magnitude (bottom panel). The black dotted and blue dashed lines in the top panel indicate the average magnitude for the whole dataset and the data used for the colour indices, respectively. In the middle panel, the blue dashed line marks the average colour index.}
    \label{colori}
    \end{figure}

The contribution of thermal radiation from the accretion disc is expected to lead to a redder-when-brighter spectral behaviour in the optical band. We verified this by performing a colour analysis. Figure \ref{colori} shows the source $R$-band light curve and the $B-R$ colour behaviour since 1995. Pre-GASP data were provided from the Abastumani, Calar Alto, Crimean, Mt.\ Maidanak, Torino\footnote{Data taken at the Torino Observatory in 1995--1996 were published in \citet{vil97} and \citet{rai98a}, but for the present paper we reprocessed all the frames with both aperture photometry and Gaussian fitting of the PSF; in particular, we carefully re-analysed the observations of 1995 June 27--28 that led to the detection of the optical outburst announced by \citet{bos95}.}, and St.\ Petersburg observatories. 
The historical light curve in the top panel shows that the source was very active in 1995--1998 and subsequently faded to $R > 16$ until 2007, when it reached $R \sim 15.5$.
Other flares peaking at $R \sim 15.5$--16 occurred in 2009 and at the beginning of 2011, while during the big outburst of mid 2011 the source returned to its previous brightness levels of 1995 and 1997 ($R \la 15$).

We derived $B-R$ colour indices by coupling the highest-quality $R$ and $B$ data (with errors smaller than 0.05 and 0.1 mag, respectively) acquired by the same telescope within 20 minutes. We obtained 372 indices, with a mean time separation between $B$ and $R$ exposures of about 6 minutes. The lack of a $B$-band light curve with sampling equivalent to the $R$-band one makes the $B-R$ versus time plot rather discontinuous. In particular, most of the colour indices were derived during faint states of the source, as illustrated by the difference between the average magnitude of the whole dataset ($R=16.42$) and that of the data used to get the colour indices ($R=16.98$). 
The minimum and maximum values of the colour index are 0.39 and 1.04, respectively, and its average value is $<B-R>=0.64$, with a standard deviation $\sigma=0.14$. 

The dependence of the colour index on brightness appears complex, with a large dispersion in faint states.
However, a general redder-when-brighter trend is recognisable for $R \ga 16$; in brighter states, the colour index may even decrease. This is reminiscent of the ``saturation effect" first noticed by \citet{vil06} for another FSRQ: 3C 454.3.
The shift from a redder-when-brighter to a bluer-when-brighter trend as the brightness increases would mark the transition from a quasar-like to a BL Lac-like spectral behaviour, the latter occurring when the synchrotron radiation dominates over the thermal emission.
The dispersion in the data points during faint states is apparently caused by at a given magnitude the source appearing to be redder during active periods than in more quiescent phases, as the comparison between the upper and middle panel of Fig.\ \ref{colori} suggests. However, we cannot rule out that at least part of this effect is due to an unverifiable photometric offset between different datasets that do not overlap in time.

Furthermore, assuming as a first approximation that the source spectrum in the optical band follows a power-law ($F_\nu \propto \nu^{-\alpha}$), we can derive the energy index $\alpha=[(B-R)-0.32]/0.41$, where we adopted Galactic extinction values of 0.048 and 0.030 mag in the $B$ and $R$ bands\footnote{From the NASA/IPAC Extragalactic Database, {\tt http://ned.ipac.caltech.edu/}}, respectively, and effective wavelengths as well as zero-mag fluxes by \citet{bes98}. The range of colour index values cited above then translates into a range of energy index values from 0.17 to 1.76. Thus, in the usual spectral energy distribution (SED) representation $\log (\nu F_\nu)$ versus $\log \nu$, the optical spectrum of 4C 38.41 is flat when $B-R$ is low, i.e.\ in general during faint states, while it becomes steep for $B-R>0.73$, which usually corresponds to bright levels.

\subsection{Rapid optical variability}

   \begin{figure}
   \centering
   \resizebox{\hsize}{!}{\includegraphics{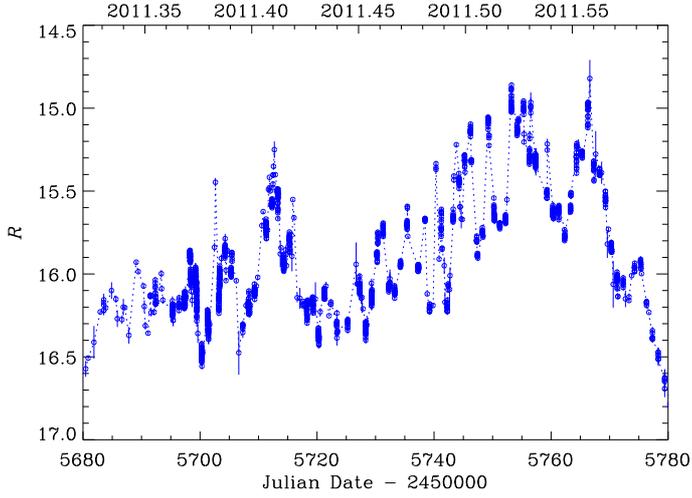}}
   \caption{The $R$-band light curve of 4C 38.41 in the most active period of 2011, showing noticeable intraday and interday variability episodes. 
The total number of data points is 2046, 1686 of which come from the Mt.\ Maidanak Observatory.}
    \label{rlc}
    \end{figure}

We analysed the $R$-band light curve presented in Fig.\ \ref{ottico} to search for the most noticeable intraday variability (IDV) episodes, involving changes of more than 0.3 mag in less than six hours. We found six nights where such rapidly occurring episodes were observed: 
on $\rm JD=2455065$, Tijarafe observations following St.\ Petersburg ones revealed a brightening of 0.32 mag in 2.82 hours (0.113 mag/hour);
on $\rm JD=2455699$, a very dense monitoring at the Mt.\ Maidanak Observatory showed a fading of 0.31 mag in 4.15 hours (0.075 mag/hour); 
on $\rm JD=2455702$, observations at the KVA revealed a 0.39 mag brightening with respect to St.\ Petersburg observations about four hours before (0.099 mag/hour);
on $\rm JD=2455715$, we derived a 0.34 mag change in 3.41 hours from the data of Goddard and Lowell (0.100 mag/hour);
on $\rm JD=2455756$, intense monitoring at the Mt.\ Maidanak and Abastumani observatories revealed variations of 0.34 mag in 4.7 hours (0.072 mag/hour);
finally, on $\rm JD=2455759$  a very rapid brightening of 0.31 mag in 0.74 hours was inferred by comparing Crimean with Mt.\ Maidanak data, which implies an impressive rate of 0.42 mag/hour. A careful check of the corresponding frames and photometry confirmed the results, so we conclude that this extreme variability episode is most likely genuine. 
A similarly rapid variability episode was reported by \citet{rai08c} for another quasar-type blazar, 3C 454.3, in January 2008 (0.40 mag/hour), and an even more extreme episode was discussed by \citet{rai08b} for the same object, which brightened by 1.1 mag in 1.5 hours and then dimmed by 1.2 mag in one hour in December 2007. 

Moreover, 4C 38.41 exhibited extraordinary optical activity in 2011, with flux variations of several tenths of a magnitude on a day timescale, as shown by Fig.\ \ref{rlc}.
The most noticeable one was a dimming of 0.77 mag in 13.81 hours observed on $\rm JD=2455702$--2455703, while variations $\ga 1$ mag were detected in 52 hours on $\rm JD=2455700$--2455702 and in 72 hours on $\rm JD=2455805$--2455808.

\section{Optical polarimetry, spectropolarimetry, and spectrophotometry}
\label{sec_stw}

Polarimetric data for the present paper were provided by the Steward, Crimean, Lowell, Calar Alto, and St.\ Petersburg observatories.
The Steward Observatory also supplied spectropolarimetry and spectrophotometry data.

   \begin{figure}
   \centering
   \resizebox{\hsize}{!}{\includegraphics{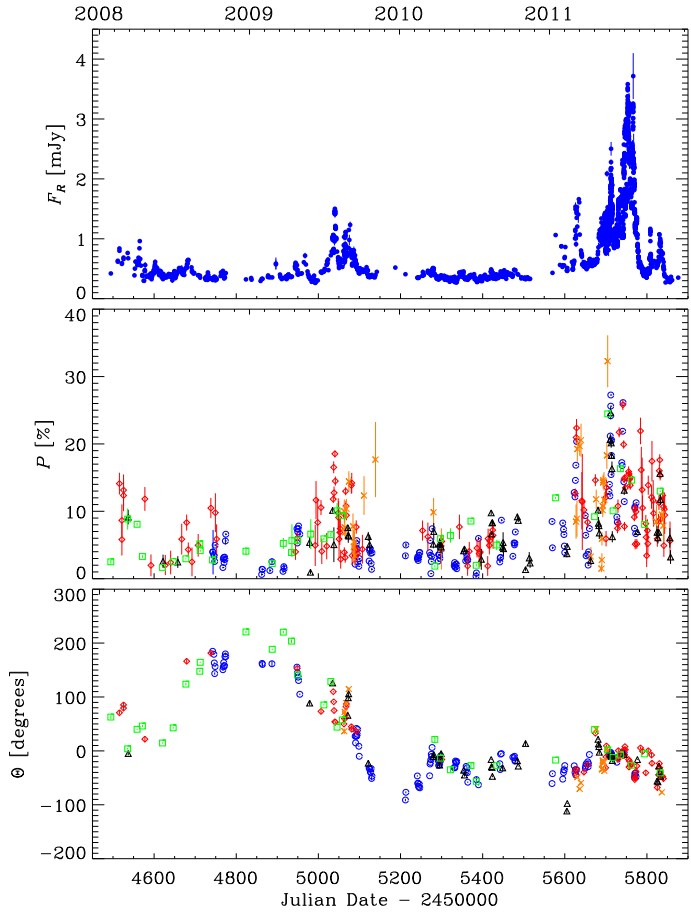}}
   \caption{Optical flux densities in the $R$ band (top panel), percentage of polarised flux (middle panel), and polarisation position angle (bottom).
Data are from the following observatories:  Steward (blue circles), Crimean (red diamonds), Lowell (black triangles), Calar Alto (green squares), and St.\ Petersburg (orange crosses). The $\pm 180\degr n$ ambiguity in $\Theta$ is solved by using only data for which $P/\sigma_P>5$ and then requiring that the change across contiguous epochs be $\le 90\degr$.}
    \label{pola}
    \end{figure}

Figure \ref{pola} shows the behaviour of the observed degree of optical polarisation $P$, as well as that of the position angle of the polarisation vector $\Theta$ as a function of time. To solve the $\pm 180\degr n$ ambiguity in $\Theta$, we consider only data for which $P/\sigma_P>5$ and then require that the change across contiguous epochs be $\le 90\degr$. Visual inspection of the figure reveals that in general periods of high flux correspond to periods of high polarisation degree, so that there is some correlation between $P$ and brightness, which is investigated further below.
For the polarisation position angle, there appears to be a smooth rotation of about 200\degr\ in 0.6 years, from 2008.4 to 2009.0, followed by an ever increasing rotation of about 270\degr\ in 0.6 years, from 2009.2 to 2009.8. Thereafter, $\Theta$ shows only smaller fluctuations between $\sim -100\degr$ and 40\degr in 2010 and 2011.
Hence, the 2009 outburst occurred during a phase of noticeable change in $\Theta$, in contrast to the 2011 one.

\subsection{Emission lines}
\label{sec_lines}

   \begin{figure}
   \centering
   \resizebox{\hsize}{!}{\includegraphics{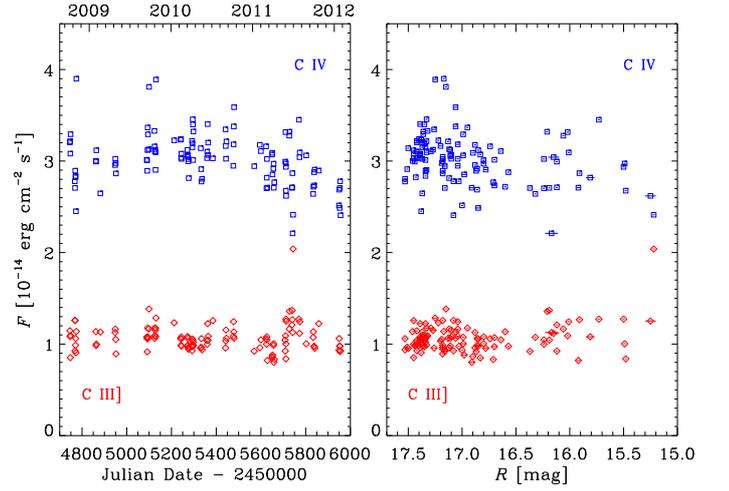}}
   \caption{The flux in the observed frame of the \ion{C}{IV} (blue squares) and \ion{C}{III]} (red diamonds) emission lines as a function of time (left panel) and source brightness (right panel). During the period of monitoring, there is little evidence of significant emission-line variability in 4C 38.41. Data are derived from the ongoing monitoring programme at Steward Observatory.}
    \label{lines}
    \end{figure}

With a redshift of $z=1.81$, two prominent broad emission-lines fall within the 4000--7550 \AA\ spectral window that is sampled by the Steward Observatory spectrophotometry of 4C 38.41: \ion{C}{IV}$\lambda$1550 and \ion{C}{III]}$\lambda$1909.  
In Fig.\ \ref{lines}, we plot the emission-line fluxes measured on 116 nights from 2008 October to 2012 January against both time and the optical brightness of 4C 38.41. 
The same trends for the line fluxes are observed as seen for the Balmer lines in PKS 1222+216 \citep{smi11} and 3C 454.3 \citep{rai08c}.
That is, the data are consistent with the line fluxes being constant, the scatter in the measurements seen within an observing campaign (about a week long) being roughly the same magnitude as during the entire monitoring period.  
Likewise, the substantial variations in the equivalent widths of the lines in 4C 38.41 (ranging from $-83$ to $-11$ \AA\ for \ion{C}{IV}) are consistent with only the continuum having varied since 2008 August.
The size of the broad-line region in quasars \citep{kas00} and the added delay in variability caused by the redshift of the object make significant line variability impossible on timescales as short as a week, so the scatter in the data is a reflection of the systematic uncertainties in the spectrophotometric measurements.  
The average \ion{C}{IV} line flux is measured to be $(3.02 \pm 0.28) \times 10^{-14} \rm \, erg \, cm^{-2} \, s^{-1}$. 
For \ion{C}{III]}, the average flux is $(1.07 \pm 0.15) \times 10^{-14} \rm \, erg \, cm^{-2} \, s^{-1}$.  
The line fluxes were measured by fitting a single Gaussian to the line profiles. 
The Gaussian fits yield an average full width at half maximum (FWHM) of $(4800 \pm 300) \rm \, km \, s^{-1}$ for \ion{C}{IV} and $(6500 \pm 1200) \rm \, km \, s^{-1}$ for \ion{C}{III]}. The large uncertainty in the measured width of \ion{C}{III]} is due to its relatively small equivalent width, especially when 4C 38.41 is bright, and its larger FWHM relative to \ion{C}{IV} is likely caused by the inclusion of \ion{Al}{III}$\lambda$1858 within the fitted line profile.

\subsection{Continuum flux and colour}
\label{sec_colour}

   \begin{figure}
   \centering
   \resizebox{\hsize}{!}{\includegraphics{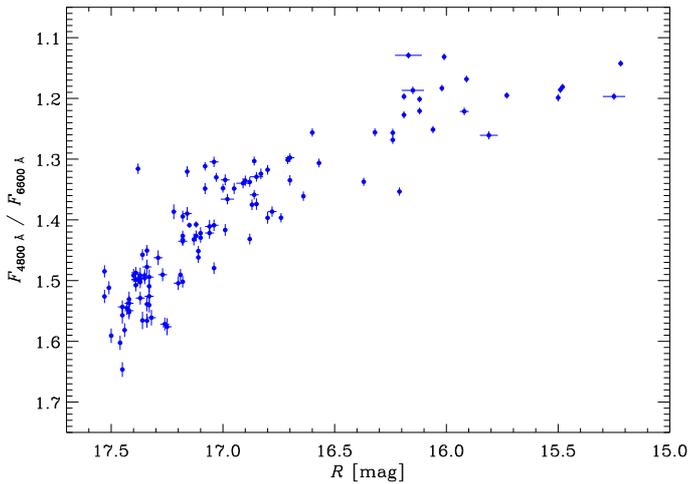}}
   \caption{The continuum colour of 4C 38.41 as a function of the source brightness. The continuum colour is approximated by the flux ratio of the two 400 \AA-wide bins centred on 4800 \AA\ and 6600 \AA\ in the observed reference frame. }
    \label{colour}
    \end{figure}

The Steward spectrophotometry can be used to examine how the optical continuum of 4C 38.41 varies with time and flux as shown in Fig.\ \ref{colori}, but avoiding the non-variable emission features that fall within the standard photometric-filter bandpasses.  
Figure \ref{colour} shows the colour of 4C 38.41 as a function of brightness in the $R$ band.
The colour is determined by taking the ratio of the fluxes within two 400 \AA-wide bins centred on 4800 \AA\ (``blue") and 6600 \AA\ (``red").  Although the spectra extend to 7550 \AA, using a redder measurement would include uncorrected terrestrial oxygen and water absorption features and would corrupt the determination of the continuum flux ratio.  The trend in Fig.\ \ref{colour} of the continuum becoming redder as the object brightens confirms the broad-band filter results shown in Fig.\ \ref{colori}.
Despite a range in wavelength of only 640 \AA\ between the flux bins in the rest frame of the object, the correlation between colour and brightness is pronounced, and suggests that the dominant source of continuum light is different when 4C 38.41 is bright than when the blazar is faint.

\subsection{Polarisation and flux}

   \begin{figure}
   \centering
   \resizebox{\hsize}{!}{\includegraphics{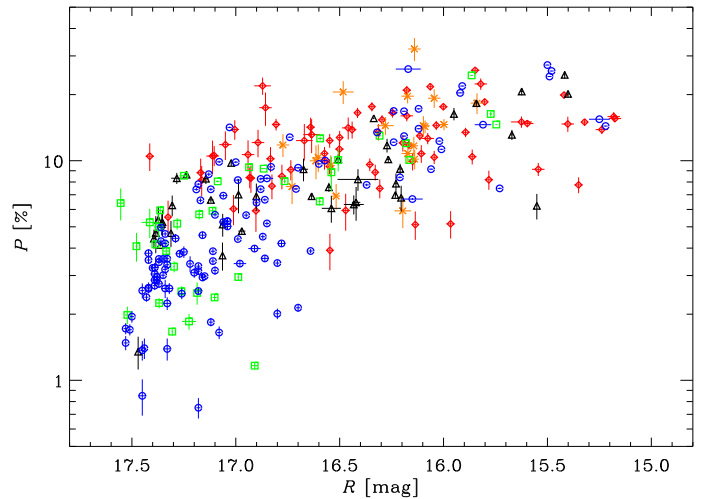}}
   \caption{The observed degree of optical polarisation plotted against the brightness of 4C 38.41 in the $R$ band. Data are from the following observatories:  Steward (blue circles), Crimean (red diamonds), Lowell (black triangles), Calar Alto (green squares), and St.\ Petersburg (orange crosses). Only data having $P/\sigma_P>5$ are included for clarity.}
    \label{polar}
    \end{figure}

\begin{figure}
   \centering
   \resizebox{\hsize}{!}{\includegraphics{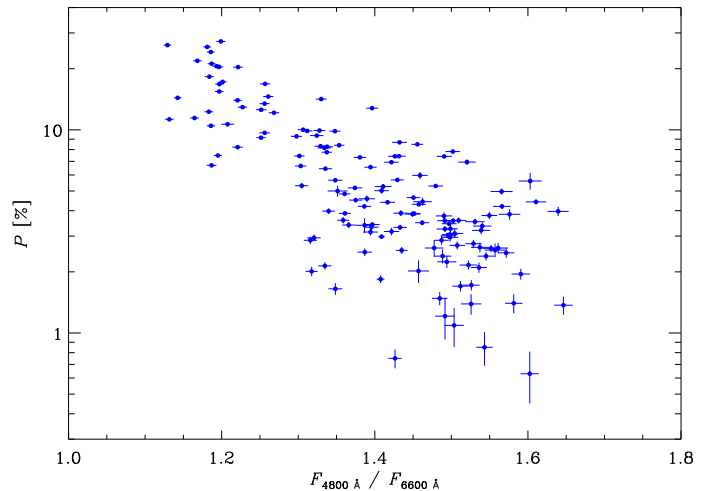}}
   \caption{The observed degree of polarisation plotted against the optical continuum colour (as defined in Fig.\ \ref{colour}) of 4C 38.41. 
The polarisation and flux ratio measurements are simultaneous, with the polarisation data being derived from the median value within a 5000--7000 \AA\ bin from the Steward Observatory spectropolarimetry.}
    \label{polflu}
    \end{figure}

A strong correlation between the degree of optical linear polarisation, $P$, and the $R$-band optical brightness is evident from the data and shown in Fig.\ \ref{polar}. Generally, 4C 38.41 is more highly polarised when bright than when faint.
Given the correlation identified between optical continuum colour and brightness in Sect.\ \ref{sec_colour}, the trend observed between $P$ and brightness translates into a strong correlation between optical colour and degree of polarisation. 
This correlation is shown in Fig.\ \ref{polflu}, with the continuum flux ratio of the 4800 \AA\ to 6600 \AA\ bins used as the measure of optical colour. 
High polarisation in 4C 38.41 is associated with a redder continuum.
This result is further evidence that the dominant source of continuum flux is different when the object is bright than when it is faint, as suggested by the observed relationship between the flux and colour.
When bright, 4C 38.41 is generally optically redder and more highly polarised.

In addition to the degree of polarisation, we searched for correlations between the polarisation position angle and optical flux, colour, and $P$. No apparent correlations were found.

\citet{jor11} found that at least from June to November 2009 there was good agreement between the position angle of optical polarisation and that of polarisation in the 43 GHz core, and between the general behaviour of $P$ in the two bands, except for the time when, according to the VLBA images, a superluminal knot was passing through the core. At this time, the optical polarization peaked,
while $P$ in the core was minimal and the optical position angle and the position angle in the core differ by $\sim 90 \degr$. This implies that the core was optically thick, most likely due to passage of the knot, which is consistent with the knot kinematics.

\subsection{Wavelength dependence of the optical polarisation}
 
   \begin{figure}
   \centering
   \resizebox{\hsize}{!}{\includegraphics{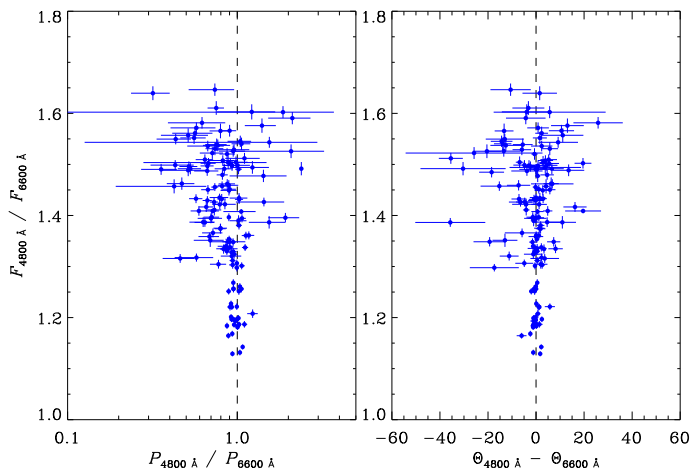}}
   \caption{Left panel: relationship between the ratio of observed polarisation in the blue to red bins and colour, defined as in Fig.\ \ref{colour}.
Right panel: relationship between the difference between the polarisation position angles in the blue and red bins, and colour.}
    \label{wavedep}
    \end{figure}

The striking correlations observed in 4C 38.41 between the optical flux, colour, and polarisation have been noted in other strong emission-line blazars \citep[see e.g.][]{sit85,smi86,smi11} and have been found to be in the same sense.  That is, as objects fade, they become bluer and are generally not as highly polarised.  During optical flares, a redder and more highly polarised source of continuum flux appears to dominate the optical continuum.  The short (daily) variability timescales and the power-law spectrum at optical wavelengths of the variable component indicate that this is the beamed synchrotron continuum from the relativistic jet in these sources.  As a flaring object fades, one can wonder whether the trends seen between flux, colour, and polarisation are simply a reflection of the evolution of the jet, or if fainter emission contributions, not directly associated with the beamed continuum, are becoming more dominant. A key to understanding these trends has been investigations of how the linear polarisation varies with wavelength.  We used 144 available spectropolarimetric observations of 4C 38.41 obtained between 2008 October and the end of 2012 January to examine the optical spectra of $P$, $\Theta$, and the polarised flux. 

Figure \ref{wavedep} shows the ratio of the fluxes in the blue to red bins (defined in Sect.\ \ref{sec_colour}) plotted against the ratio of the observed polarisation in the same bins.  As the blazar becomes bluer (fainter), the polarisation in the blue tends to be less than $P$ measured in the red bin.  
The strength of this wavelength dependence in $P$ generally increases as the object becomes bluer (fainter), although the low levels of polarisation and flux encountered result in relatively large uncertainties in $P_{4800 \AA}/P_{6600 \AA}$.
Figure \ref{wavedep} also shows the difference between the polarisation position angles determined in the two continuum  bins. Most observations are consistent with no wavelength dependence in $\Theta$. 
This general trend of $\Theta$ being constant in wavelength suggests that there is only a single source of polarised flux from the jet that dominates at any given time.  
If a second non-thermal source of polarised flux that has a significantly different spectral index and polarisation position angle becomes bright enough to compete with the original source, then a strong wavelength dependence in $\Theta$ can occur \citep[see e.g.][]{hol84,sit84}, although this does not seem to happen often \citep{sit85,smi87,mea90}.  
Much of the wavelength dependence may of course be hidden if the sources of polarised flux have similar spectral indices and/or polarisation position angles within the spectral region being observed.

To more clearly illustrate the spectral dependence of $P$ for an object as faint as 4C 38.41, we took a median of the 144 available polarisation spectra obtained throughout the monitoring period.  This was done by rotating each polarisation spectrum so that all of the polarised flux is in the $q$ Stokes parameter.  That is, the observed $q$ and $u$ spectra were transformed (rotated) to $q'$ and $u'$, where $u'$ averages to 0 over the entire spectrum.  The resulting $q'$ spectra were then scaled to $q' = 0.1$ (10\% polarisation) within the 6400--6800 \AA\ bin and then median Stokes spectra were obtained.  An important advantage of using $q'$ instead of $P$ is that the Stokes parameters have normal error distributions, while $P$ is biased since it is a positive, definite quantity \citep[e.g.][]{war74}.

   \begin{figure}
   \centering
   \resizebox{\hsize}{!}{\includegraphics{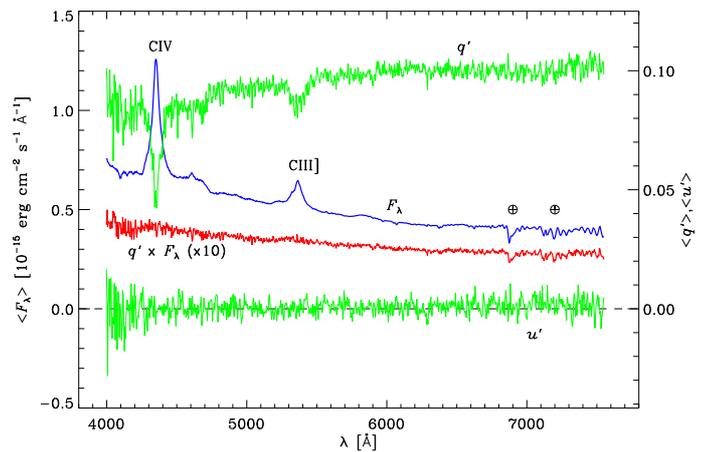}}
   \caption{The median flux (blue), polarised flux (red), and normalised $q'$ and $u'$ Stokes (green) spectra for 4C 38.41.
These spectra were compiled from 144 spectropolarimetric observations of the blazar. The polarised flux spectrum is derived by multiplying the flux spectrum by $q'$ and displays a featureless power-law as expected from a synchrotron source. The polarised flux spectrum is scaled by a factor of 10 for clarity. See the text for a full description of the $q'$ and $u'$ spectra.
Telluric absorption features are identified with $\oplus$ symbols.}
    \label{stokes}
    \end{figure}

The median $q'$ and $u'$ spectra for 4C 38.41 are displayed in Fig.\ \ref{stokes} and identify the likely cause of the general decrease in the polarisation toward the blue end of the spectrum.
In addition to a decrease in $q'$ for the continuum at $\lambda \la 6000$ \AA, there are marked decreases in the polarisation at \ion{C}{IV} and \ion{C}{III]}.  
This indicates that the broad emission lines are not highly polarised, if at all.  
Indeed, the $q'$ spectrum appears to be very similar to an inverted spectrum of a QSO down to the blended \ion{He}{II}+\ion{O}{III]} emission feature close to the red wing of \ion{C}{IV} and suggests that the observed wavelength dependence is caused by the dilution of the polarised flux by an unpolarised spectrum that is similar to that of an optically selected QSO.
The flat, featureless spectrum of $u'$ supports the assertion that was made from Fig.\ \ref{wavedep} that any wavelength dependence in the optical polarisation position angle is transitory and typically not strong.
  
Figure \ref{stokes} also displays the median flux ($F_{\lambda}$) and polarised flux ($q' \times F_{\lambda}$) spectra of 4C 38.41.
Here, the nature of the source of polarised flux becomes evident.
The median polarised spectrum is featureless and well fit by a power-law with a spectral index $-1.25$, which is generally redder than the continuum of the median total flux spectrum.
The spectral index of the polarised continuum is identical to that of the synchrotron continuum under the assumption that the intrinsic jet polarisation ($P_0$) is independent of wavelength over this spectral region.  
Since the synchrotron power-law is redder than the QSO-like emission components (broad emission lines + blue continuum), $P$ decreases into the blue. 
The amount of dilution and the strength of the wavelength dependence of the polarisation are both a function of the relative brightness between the unpolarised, non-varying (at least on timescales measured in months to years) emission from the QSO component and the highly variable non-thermal emission from the relativistic jet and its spectral index.
This model not only successfully explains the wavelength dependence of the observed polarisation, but also explains the correlations seen in 4C 38.41 between continuum flux, colour, and level of polarisation because the bluer QSO component contributes more strongly to the optical spectrum when the object is faint.
The same general model has also been shown to apply to lower-redshift blazars with strong emission lines, such as 3C 345 \citep{smi86}, PKS 1546+027 \citep{smi94}, and PKS 1222+216 \citep{smi11}.
The results for 4C 38.41 extend the evidence of the two-component nature of these types of blazars into the ultraviolet down to \ion{C}{IV}$\lambda$1550.

With the polarisation and flux information in hand, it is impossible to distinguish the optical spectrum of the object into polarised power-law synchrotron and unpolarised components since it is impossible to break the degeneracy between the intrinsic polarisation ($P_0$; assumed to be constant with wavelength) and its brightness unless the SED of unpolarised component is assumed.
The problem, however, is reasonably constrained as the unpolarised component cannot be so bright that $P_0$ is driven to excessively high values. The intrinsic polarisation certainly must be $<100$\%, but $P_0<40$--45\%, which is the observed maximum for blazars \citep[see e.g.][]{imp82,mea90}, is likely to be the upper limit.  
In addition, given the longer timescales observed for variations in AGNs that are not blazars, the unpolarised component cannot be brighter than the optical photometric minimum observed in the light curves shown in Fig.\ \ref{ottico} and Fig.\ \ref{colori} (roughly $R \sim 17.8$). 

Two methods have been used to break the $P_0$-brightness degeneracy to estimate the true polarisation of the optical synchrotron continuum throughout the monitoring period of 4C 38.41. 
First, the brightness of the unpolarised QSO component can be estimated from the strengths of the \ion{C}{IV} and \ion{C}{III]} emission lines based on the line-to-continuum flux ratios of a standard template spectrum for radio-quiet QSOs \citep[e.g.][]{fra91}.  
For \ion{C}{IV}, this yields an $R$-band magnitude for the QSO component of 17.3, and for \ion{C}{III]}, $R \sim 18.0$.  The discrepancy is caused by the \ion{C}{IV}/\ion{C}{III]} line ratio in 4C 38.41  differing significantly from the chosen template QSO, which cannot be explained by reddening.  Alternatively, if each spectropolarimetric observation of 4C 38.41 is fit with a combination of the QSO template and polarised power-law, ignoring the line fluxes and leaving the non-thermal spectral index, QSO brightness, and $P_0$ as free parameters, we find an average $R$ magnitude of 17.85 for the QSO.  
The median spectral index of the polarised power-law from these 106 fits is $-1.6$, as compared to $-1.25$ found from the median spectrum of $q'$.

   \begin{figure}
   \centering
   \resizebox{\hsize}{!}{\includegraphics{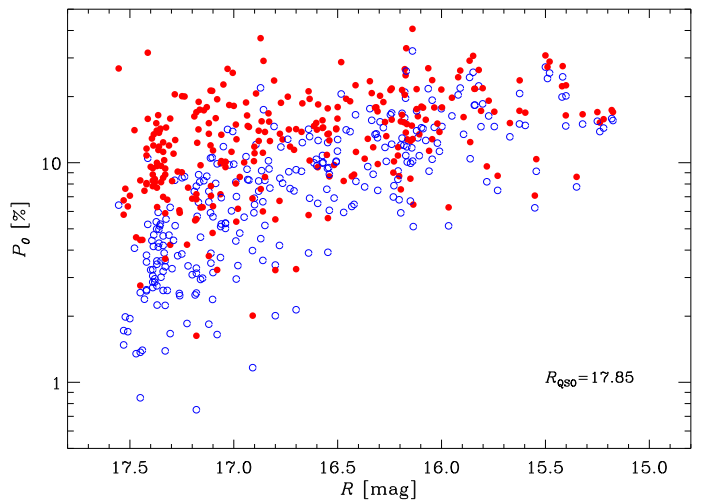}}
   \caption{The intrinsic polarisation of the optical synchrotron emission $P_0$ (red filled symbols) as a function of the brightness of 4C 38.41.
For comparison, the observed polarisation measurements (Fig.\ \ref{polar}) are also shown (blue empty symbols). 
Correcting $P$ for unpolarised emission from a QSO-like component with $R = 17.85$ yields $P_0$ (see text).}
    \label{pola_cor}
    \end{figure}

The estimate of the QSO brightness based on the strength of \ion{C}{IV} can be ruled out, as it would make the overall $R$-band light curve difficult to explain because 4C 38.41 often becomes fainter.  
The estimates based on \ion{C}{III]} and the model fits are consistent with the observed light curve.  
The intrinsic polarisation of the synchrotron continuum is given by $P_0=P_{\rm obs} \times F_{\rm obs} /(F_{\rm obs}-F_{\rm QSO})$, where $P_{\rm obs}$ and $F_{\rm obs}$ are the observed polarisation and flux within the $R$ filter bandpass. 
In Fig.\ \ref{pola_cor}, the intrinsic polarisation is plotted against the brightness of the blazar, assuming $R = 17.85$ for the unpolarised QSO component. Similar results are obtained if $R=18$ is assumed, but with smaller corrections being required between the observed and intrinsic polarisations.
The correction of $P$ for the non-beamed nuclear emission greatly lessens the correlation between $P$ and optical brightness displayed in Fig.\ \ref{polar}. The correlation is not completely destroyed as there are no cases of $P \la 5$\% being observed when 4C 38.41 is in a major optical outburst, but $P_0$ approaching 20--30\% can occur when the object is near its faintest flux levels during the monitoring period.

\section{Long-term observations at radio and millimetre wavelengths}

   \begin{figure*}
   \centering
   \includegraphics{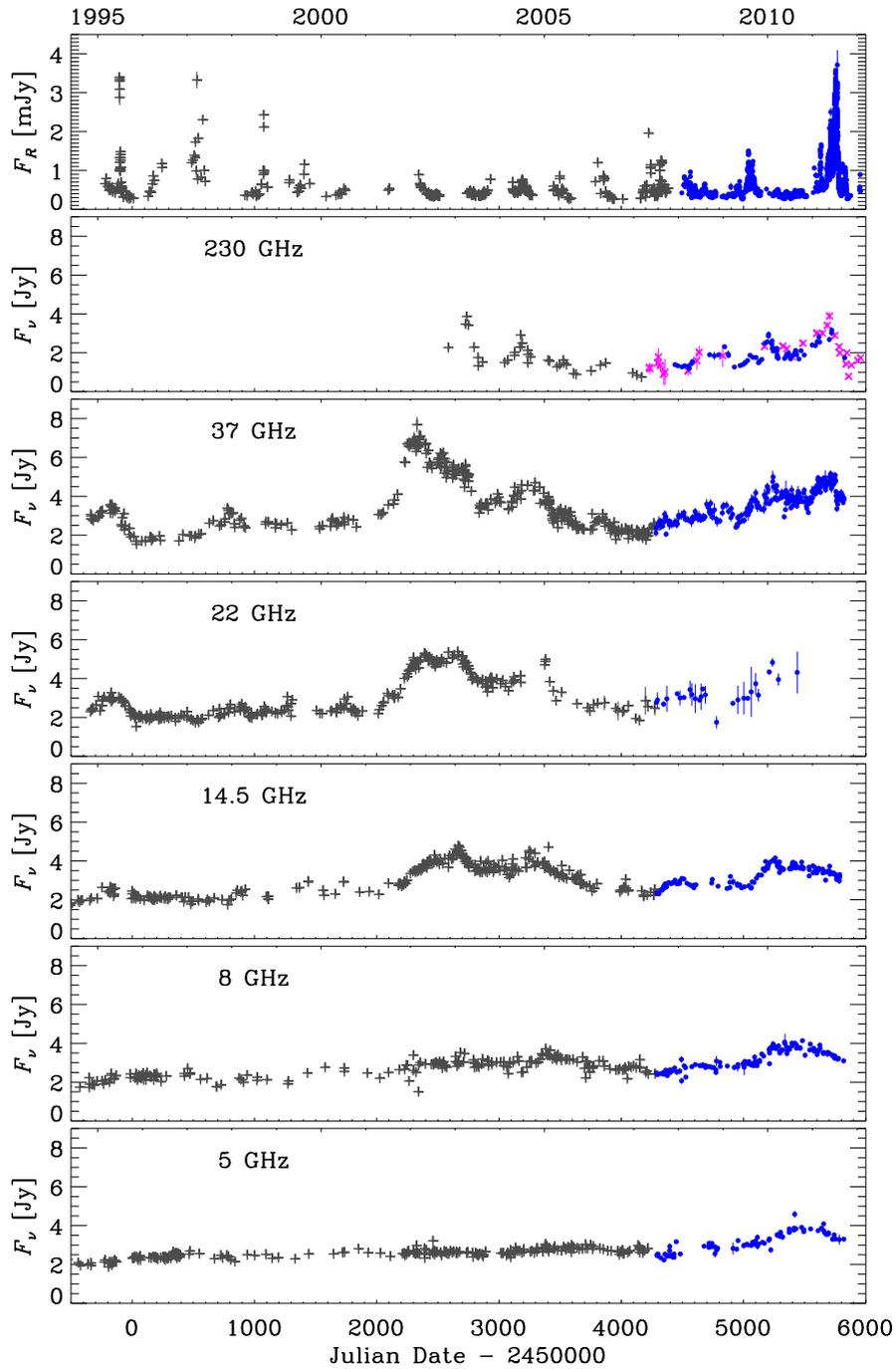}
   \caption{$R$-band optical flux densities (top) compared to the radio light curves at different frequencies in 1994--2011. Blue dots refer to data collected in the ambit of the GASP project, complemented by 230 GHz data from the IRAM 30 m telescope (purple crosses). Black plus sign symbols are historical data taken from the participating observatories' archives and from the literature (see text for details).}
    \label{radop}
    \end{figure*}

We analysed radio data provided to the GASP project by
the Submillimeter Array (SMA, 230, 275, and 345 GHz\footnote{These data were obtained as part of the normal monitoring programme initiated by the SMA \citep[see][]{gur07} as well as part of a dedicated programme by Ann Wehrle.}) and by the radio telescopes of Medicina (5, 8, and 22 GHz), Mets\"ahovi (37 GHz), Noto (38 and 43 GHz), and UMRAO (4.8, 8.0, and 14.5 GHz).
Additional data at 86 and 230 GHz were supplied for this paper by the IRAM 30\,m telescope\footnote{IRAM 30\,m data were acquired as part of the POLAMI (Polarimetric AGN Monitoring with the IRAM-30\,m-Telescope) and MAPI (Monitoring AGN with Polarimetry at the IRAM-30\,m-Telescope) programmes. Data reduction was performed following the procedures described in \citet{agu06,agu10}.}.
We also collected archival data, which were supplied by the SMA (230 GHz), Medicina (5, 8, and 22 GHz), Mets\"ahovi (37 GHz), Noto (5, 8, 22 and 43 GHz), and UMRAO (4.8, 8.0, and 14.5 GHz) telescopes, as well as 22, 37, and 87 GHz data acquired with the Mets\"ahovi and Crimean radio antennas and published in \citet{sal87} and \citet{ter92,ter98,ter04,ter05}.
The long-term radio light curves of 4C 38.41 at the best-sampled wavelengths are displayed in Fig.\ \ref{radop}, starting from $\rm JD=2449500$. The radio flux densities are compared with the optical $R$-band ones. 

\begin{figure}
   \centering
   \resizebox{\hsize}{!}{\includegraphics{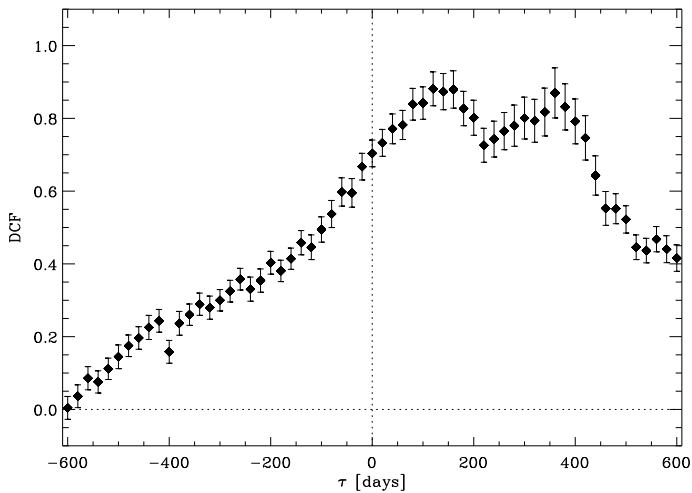}}
   \caption{Discrete correlation function between the 37 GHz and 14.5 GHz light curves shown in Fig.\ \ref{radop}.}
    \label{dcf_radio}
    \end{figure}

The variability maximum amplitude ($F_{\rm max} / F_{\rm min}$) decreases with decreasing frequency (with the exception of the 230 GHz light curve, which lacks data before 2002), suggesting that the corresponding jet emitting regions become larger. The shape of the major event in 2001--2003, clearly visible at 37, 22, and 14.5 GHz, changes with wavelength: it shows two peaks of similar intensity at 22 GHz, while at 37 (14.5) GHz the first (second) peak prevails. It is hard to say whether we are seeing here the evolution of a single event, or rather the effect of two overlapping events.
Cross-correlation of the 37 and 14.5 GHz light curves by means of the discrete correlation function \citep[DCF;][]{ede88,huf92,pet01} shows a first peak at a time lag of 120 days, and another peak at 360 days (see Fig.\ \ref{dcf_radio}), indicating a time delay of several months of longer-wavelength flux variations. This is a common feature in blazars and suggests that lower-frequency photons are produced downstream in the jet from the higher-frequency ones.

The comparison between the optical and radio-mm light curves (Fig.\ \ref{radop}) reveals a lack of general correlation: the big outbursts that occurred at high radio frequencies in 2001--2003 show no optical counterpart. Moreover, the big optical outbursts of 1995 and 1997 do not seem to have been followed by major radio events on a few-month timescale. The possibility remains that we either missed some important optical flare or that the radio-optical correlation occurs on much longer timescales than expected for this kind of objects.
While we need to wait to see whether a big radio outburst will follow the 2011 optical event, we note that the systematically higher radio flux in 2009--2011, unusually at all wavelengths, may be related to the renewed optical activity of the past years. 

In the framework of a geometrical interpretation of blazar variability previously proposed for other objects \citep[see e.g.][]{vil99,vil07,vil09a,rai09,rai10,rai11}, one could assume an inhomogeneous curved jet, where the optical and radio emitting regions were more misaligned in the past. In particular, a closer alignment of the optical region with the line of sight in 1995--1998 would have made the optical radiation more relativistically beamed toward us, while in 2001--2005 the radio-emitting region would have been better aligned, with the consequent greater Doppler enhancement of the radio flux. The increased activity of the past few years at both radio and optical frequencies would then indicate that the radio-optical misalignment has decreased and that both emitting regions are seen at small viewing angles.
An analysis of the variations in the viewing angle of the parsec-scale jet will be performed in Jorstad et al.\ (2012, in preparation).

   \begin{figure}
   \centering
   \resizebox{\hsize}{!}{\includegraphics{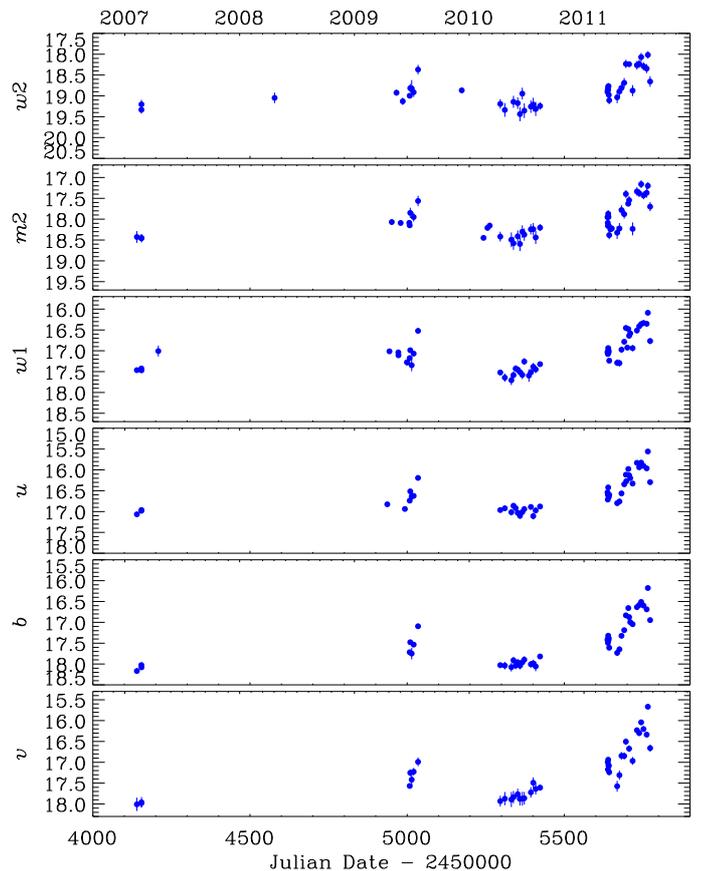}}
   \caption{Optical and UV light curves of 4C 38.41 built with {\it Swift}-UVOT data.}
    \label{uvot}
    \end{figure}

   \begin{figure}
   \centering
   \resizebox{\hsize}{!}{\includegraphics{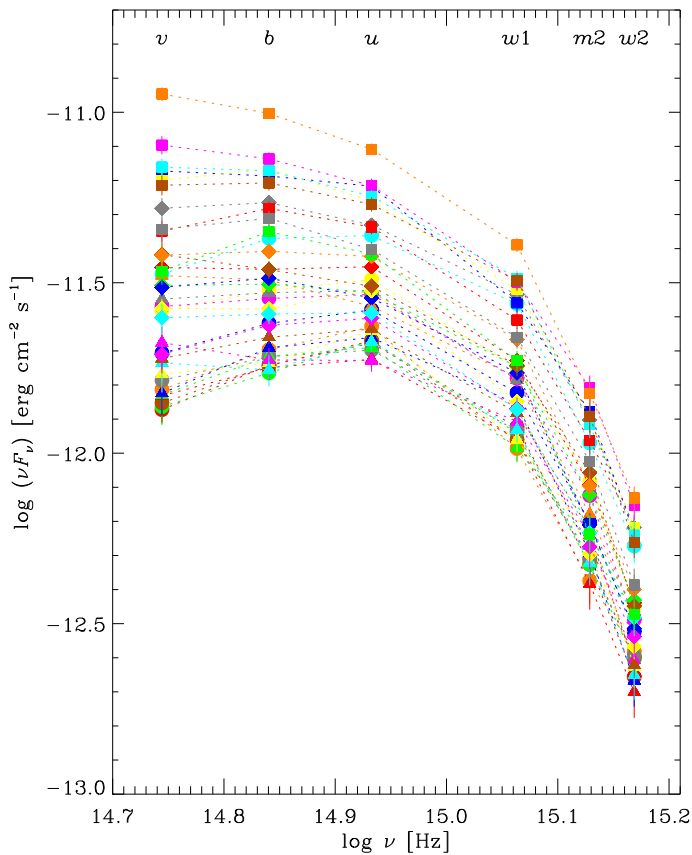}}
   \caption{Spectral energy distributions of 4C 38.41 in the optical--UV frequency range compiled with UVOT data. As the jet synchrotron emission becomes fainter, the contribution of the thermal radiation from the accretion disc emerges, peaking around the $u$ band.}
    \label{sed_uvot}
    \end{figure}

\section{{\it Swift} observations}
\label{sec_swift}
The {\it Swift} spacecraft \citep{geh04} carries three instruments: the Burst Alert Telescope (BAT; \citealt{bar05}), observing between 15 keV and 150 keV; the X-ray Telescope (XRT; \citealt{bur05}), observing between 0.3 and 10 keV; and the Ultraviolet/Optical Telescope (UVOT; \citealt{rom05}), acquiring data in the 170--600 nm range.
Up to 2011 September 30, {\it Swift} targeted 69 times 4C 38.41.

\subsection{UVOT observations}
\label{sec_uvot}
The UVOT observations were performed in the optical $v$, $b$, and $u$ bands, as well as in the UV filters $uvw1$, $uvm2$, and $uvw2$ \citep{poo08}. 
We reduced the data with the HEAsoft package version 6.10, with the 20101130 release of the {\it Swift}/UVOTA CALDB. 
Multiple exposures in the same filter at the same epoch were summed with {\tt uvotimsum}, and aperture photometry was then performed with the task {\tt uvotsource}. Source counts were extracted from a circular region with a 5 arcsec radius, while background counts were estimated in a neighbouring source-free circular region with a 10 arcsec radius.
The results are shown in Fig.\ \ref{uvot} (and in Fig.\ \ref{ottico}, where the $v$-band points have been shifted by $-0.1$ mag to match the Johnson's $V$-band data). 
The ranges of magnitude spanned in the various bands are about
2.3, 2.0, 1.6, 1.6, 1.4, and 1.4 mag going from the $v$ to the $uvw2$ filter.
Hence, the source variability decreases as the frequency increases, extending the trend already observed in the near-IR--optical band in Sect.\ \ref{sec_opt}.

To clarify whether there are any emission contributions in the UVOT energy range, we compiled SEDs for all the 36 epochs where good observations in all the six UVOT filters were available.
We took into account the calibration updates introduced by \citet{bre11}, including in particular new $\lambda_{\rm eff}$, count-rate-to-flux conversion factors, and effective areas for the UV filters.
We calculated the (small) Galactic extinction in the various bands by convolving the \citet{car89} laws (setting $R_v=3.1$ and $A_v=0.037$\footnote{From the NASA/IPAC Extragalactic Database, {\tt http://ned.ipac.caltech.edu/}}) with the new effective areas, and used the results to obtain de-reddened flux densities. The source SEDs are shown in Fig.\ \ref{sed_uvot}.

\subsection{XRT observations}
\label{sec_xrt}

Although a few SEDs have a wavy shape, likely because of the data imprecision, most SEDs have a common pattern, with a bump peaking in the $u$ band in the faintest states turning progressively into a steeper and steeper curved spectrum as the source brightness increases.
Correction for the small flux contribution of the \ion{C}{IV} emission line (see Sect.\ \ref{sec_lines}) to the $b$ band would not change the SED shape appreciably.
This again suggests that there is a thermal contribution from the accretion disc, which peaks around the $u$ band (as expected because of the high source redshift) and emerges when the jet synchrotron contribution weakens.

\begin{figure*}
   \centering
   \includegraphics{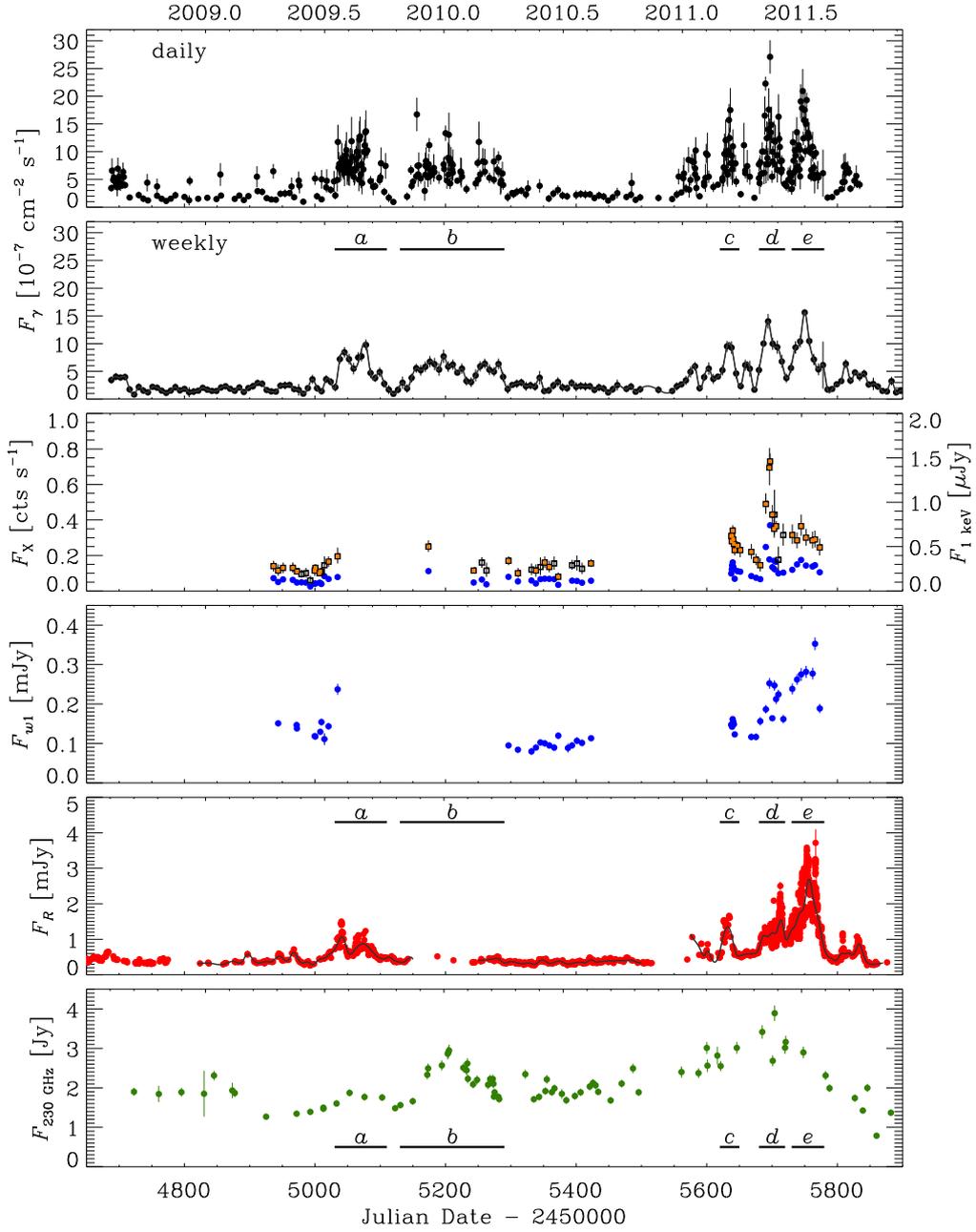}
   \caption{Light curves of 4C 38.41 at different frequencies in 2008--2011. From top to bottom: 
1) the daily $\gamma$-ray light curve from {\it Fermi}-LAT in the 100 MeV -- 300 GeV range ($10^{-7} \rm \, ph \, cm^{-2} \, s^{-1}$);
2) the corresponding weekly $\gamma$-ray light curve with its cubic spline interpolation;
3) the X-ray light curve from {\it Swift}-XRT, derived as explained in Sect.\ \ref{sec_xrt};
blue dots are counts $\rm s^{-1}$ , while black squares represent 1 keV de-absorbed fluxes ($\mu$Jy), which are filled in orange when resulting from robust spectral fits;
4) the de-absorbed {\it Swift}-UVOT flux density light curve in the $uvw1$ band (mJy);
5) the de-absorbed GASP $R$-band flux-density light curve (mJy) with the cubic spline interpolation through the 7-day binned data;
6) the millimetre radio light curve at 230 GHz (Jy).
}
    \label{multi}
    \end{figure*}

We processed the XRT event files acquired in pointing mode, using the HEASoft package version 6.11 with the calibration files 20110915\footnote{\tt http://heasarc.gsfc.nasa.gov/docs/heasarc/caldb/\\Swift/}.
We considered the observations with exposure times longer than 5 minutes, including 65 observations in photon-counting (PC) mode.
The task {\tt xrtpipeline} was run with standard filtering and screening criteria, in particular selecting event grades 0--12. 
Source counts were extracted from a 25 pixel circular region ($\sim 60$ arcsec) centred on the source, and background counts were derived from a surrounding annular region with radii of 100 and 150 pixels. The count rate was always lower than 0.5 counts/s, so no correction for pile-up was needed.
The {\tt xrtmkarf} task was used to generate ancillary response files (ARF), which account for different extraction regions, vignetting, and PSF corrections. 
The X-ray light curve (net count rate and de-absorbed 1 keV flux density) is shown in Fig.\ \ref{multi} and is discussed in the next section.

We performed spectral analysis with the {\tt xspec} package.
After ``grouping" the data with the task {\tt grppha} in order to have a minimum of 20 counts per energy channel to apply the $\chi^2$ statistics, only $\sim 20$\% of the spectra resulted in more than 10 channels. 
We therefore used the Cash statistics, which extends the $\chi^2$ statistics to the case of a low number of counts \citep{cas79,nou89}.
We applied an absorbed power-law model, where absorption is modelled according to \citet{wil00} and the hydrogen column is fixed to the LAB\footnote{\citet{kal05}; {\tt http://www.astro.uni-bonn.de}} Galactic value $N_{\rm H}=0.111 \times 10^{21} \, \rm cm^{-2}$. We tested the results of the Cash statistics with the {\tt goodness} command, which performs Monte Carlo spectral simulations based on the model and gives the percentage of them that have fit statistics smaller in value than that of the data.
Moreover, we compared the results of the Cash method with those of the $\chi^2$ statistics, when the latter was applicable. 
We obtained robust spectral fits (excluding the cases in the tails of the goodness distribution) 
in about 70\% of cases. These cases indicate a hard spectrum, with the photon index $\Gamma$ ranging from 1.31 to 1.87, with a mean value of 1.62. The lowest values correspond to the highest fluxes, in agreement with the ``harder-when-brighter" trend often observed in blazars \citep[e.g.][and references therein]{dam11}.

\begin{figure}
   \centering
   \resizebox{\hsize}{!}{\includegraphics[angle=-90]{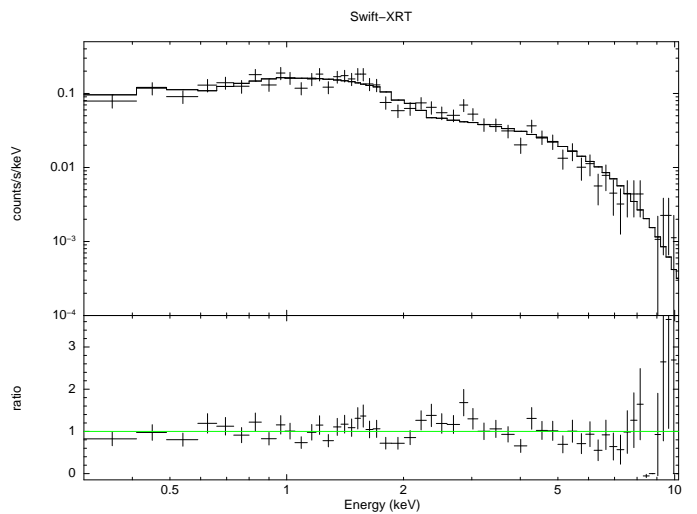}}
   \caption{The {\it Swift}-XRT spectrum of 4C 38.41 on 2011 May 15. Data are fitted with a power-law with fixed Galactic absorption. The lower panel shows the ratio of the data to the folded model.}
    \label{qqq}
    \end{figure}

In particular, the data acquired on 2011 May 15 ($\rm JD=2455697$), corresponding to the X-ray peak visible in Fig.\ \ref{multi}, are best-fitted by a power-law with $\Gamma=1.31 \pm 0.07$, 1 keV de-absorbed flux density $F_{\rm 1 \, keV}=(1.46 \pm 0.09) \, \rm \mu Jy$, and observed\footnote{The unabsorbed flux is only 1\% higher.} flux between 0.3 keV and 10 keV of $(2.24 \pm 0.14) \times 10^{-11} \rm \, erg \, cm^{-2} \, s^{-1}$. The result of the goodness test is 54\%, which assures us that the observed spectrum is well-reproduced by the model. The corresponding spectrum and the folded model is shown in Fig.\ \ref{qqq}.

\section{Cross-correlations from radio to $\gamma$-ray frequencies}
\label{sec_multi}

Figure \ref{multi} compares 4C 38.41 light curves at different frequencies in the period 2008--2011, i.e. from the beginning of the {\it Fermi} $\gamma$-ray observations. From top to bottom, one can see:
1) the daily photon flux in the 100 MeV -- 300 GeV energy range; 
2) the corresponding weekly $\gamma$-ray light curve\footnote{The daily and weekly $\gamma$-ray light curves were downloaded from the {\it Fermi}-LAT monitored source site ({\tt http://Fermi.gsfc.nasa.gov/ssc/data/access/lat}), but their preliminary nature does not affect the results of our analysis. The comparison with an independent analysis available at the ISDC {\tt http://www.isdc.unige.ch/heavens/} indeed reveals a fair agreement.};
3) the X-ray light curve from {\it Swift}-XRT (net count rate, as well as de-absorbed 1 keV flux densities derived as explained in Sect.\ \ref{sec_xrt});
4) the de-absorbed UV flux densities in the $uvw1$ band (mJy);
5) the $R$-band de-absorbed flux densities (mJy);
6) the millimetre radio light-curve at 230 GHz (Jy).

\begin{figure}
   \centering
   \resizebox{\hsize}{!}{\includegraphics{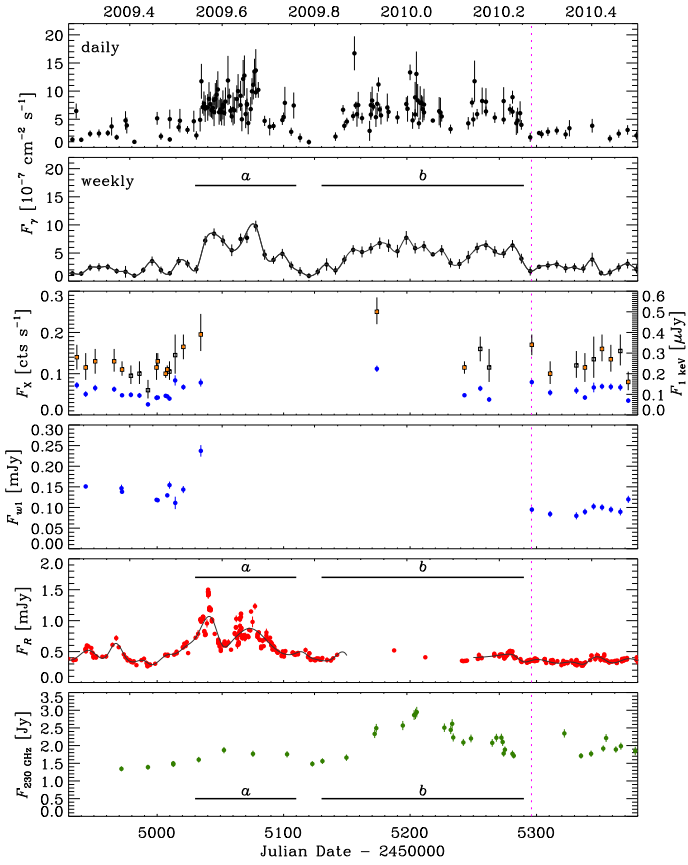}}
   \caption{An enlargement of Fig.\ \ref{multi} in 2009--2010.
The pink vertical line marks the epoch of one of the SEDs shown in Fig.\ \ref{sed}.}
    \label{multi1}
    \end{figure}

\begin{figure}
   \centering
   \resizebox{\hsize}{!}{\includegraphics{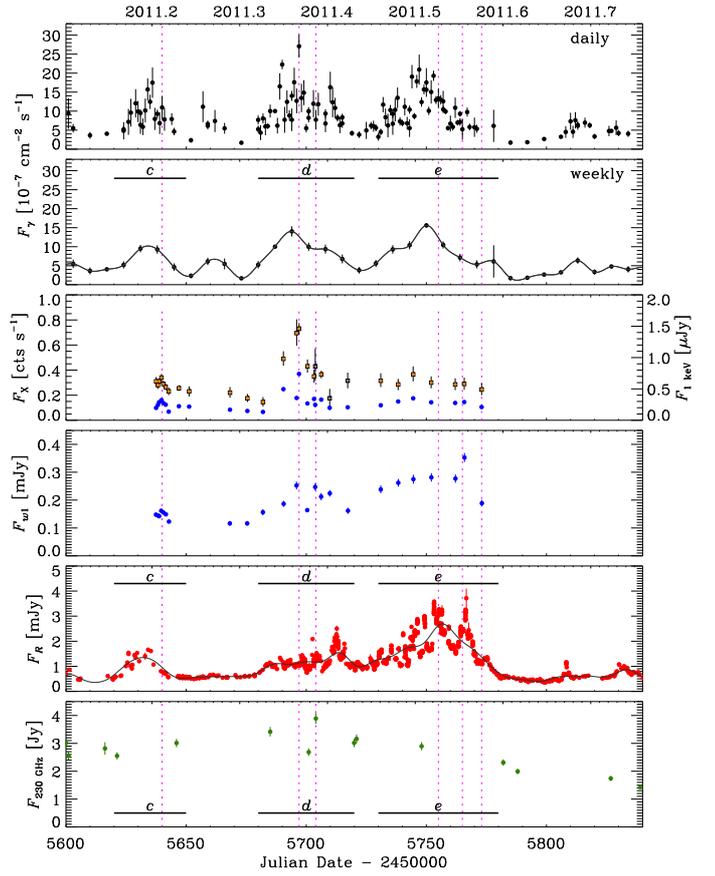}}
   \caption{An enlargement of Fig.\ \ref{multi} in 2011.
The pink vertical lines mark the epochs of six of the SEDs shown in Fig.\ \ref{sed}.}
    \label{multi2}
    \end{figure}

A zoomed image of the most interesting periods in 2009--2010 and 2011 is displayed in Figs.\ \ref{multi1} and \ref{multi2}, respectively.

To make the comparison among bands easier, we highlighted five periods, corresponding to flaring states in $\gamma$-rays. 
Moreover, we also plotted cubic spline interpolations through the weekly binned $\gamma$-ray and optical data, in the latter case to emphasize the smoothed trend of the flux, beyond the fast variability.
We note the following features:
\begin{itemize}
\item In $\gamma$-rays and in the optical band, where the sampling is dense enough, all events appear as a succession of several short-term flares. 
\item  All events (more precisely: the envelope over the substructure) show a similar duration in $\gamma$-rays and in the optical band (apart from event $b$ that does not have enough optical coverage).
\item At all frequencies, the maximum flux during the 2011 flare is greater than the maximum flux during the 2009 and 2010 flares.
\item Flare $a$ is double-peaked in both $\gamma$-rays and the optical. However, the second $\gamma$ peak is higher than the first one, while in the optical the reverse is true. A little bump is also visible at 230 GHz.
\item Flare $b$ occurred during solar conjunction so that the optical coverage is almost null. However, the few contemporaneous data do not reveal an increase in the optical flux. Interestingly, the millimetre light curve shows a bump simultaneous with the $\gamma$ event.
\item The shape of flare $c$ is similar at $\gamma$-ray and optical frequencies.
\item In flare $d$, the most active phase in $\gamma$-rays and X-rays precedes that in the optical band and the mm flux reaches its maximum value.
\item A shift between the $\gamma$-ray and optical smoothed-trend flux maxima also occurs in flare $e$, where furthermore the last optical spike has no relevant counterpart at $\gamma$-ray energies.
\end{itemize}

In conclusion, the correlation between the source emission in $\gamma$-rays and optical (and maybe millimetre) frequencies is clear, because flares are observed to occur in the same periods and their strength increases from 2009--2010 to 2011 at both frequencies. 
This would suggest that the jet regions where photons at these frequencies are produced either coincide or are very close. The more coarsely sampled X-ray and UV light curves indicate some correlation with the $\gamma$-ray and optical ones, so that the X-ray and (the non-thermal part of) the UV radiation may also come from the same jet zone.

\begin{figure}
   \centering
   \resizebox{\hsize}{!}{\includegraphics{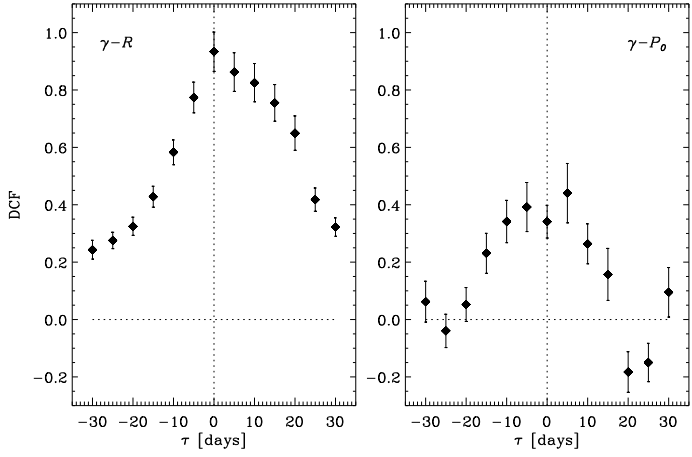}}
   \caption{Discrete correlation function between the daily $\gamma$-ray flux and $R$-band flux density light curves shown in Fig.\ \ref{multi} (left) and between the daily $\gamma$-ray flux and the QSO-corrected polarisation $P_0$ (right, see Sect.\ \ref{sec_stw}).}
    \label{dcf_gogp}
    \end{figure}

\begin{figure}
   \centering
   \resizebox{\hsize}{!}{\includegraphics{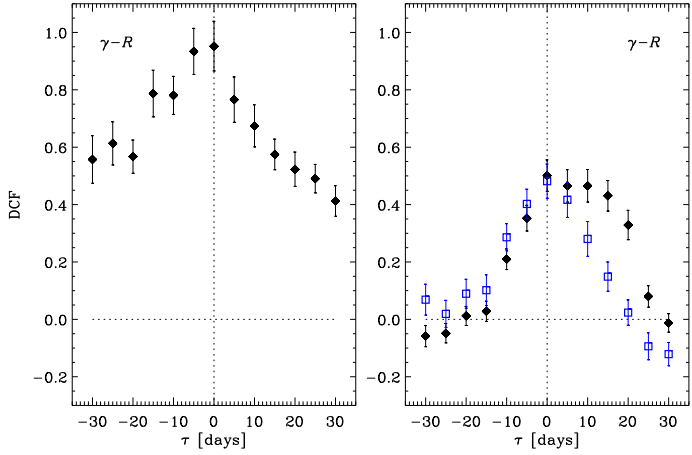}}
   \caption{Discrete correlation function between the daily $\gamma$-ray flux and $R$-band flux density light curves shown in Fig.\ \ref{multi} before (left) and after (right) $\rm JD=2455450$ for a comparison with the \citet{jor11} results (see the text). The blue squares represent the case where the light curves are stopped at $\rm JD=2455763$ to avoid the influence of the optical flare without any significant $\gamma$-ray counterpart at $\rm JD \sim 2455767$.}
    \label{dcf_go}
    \end{figure}

However, the $\gamma$-optical correlation is complex. The shape and relative brightness of the flares differ indeed considerably.
Cross-correlation of the $\gamma$-ray daily flux with the $R$-band flux by means of the DCF yields a strong bump ($\rm DCF_{\rm peak}=0.93$, see Fig.\ \ref{dcf_gogp}) peaking at zero time lag, but the centroid of the bump, which provides a more robust estimate of the time lag \citep{pet01}, appears shifted toward positive time-lags. Hence, the correlation between the $\gamma$-ray and optical fluxes is strong, and there is possibly a delay in the optical variations after those in the $\gamma$-rays, but the broadness of the DCF bump would suggest that this delay may be variable.
Moreover, this result appears in contrast to the findings of \citet{jor11}, who found a strong correlation between the optical and $\gamma$ flux variations in 2009--2010, with the $\gamma$ variations lagging by $5 \pm 3$ days. 
To investigate in greater detail the matter, we thus restrict the calculation of the DCF to the period investigated by \citet{jor11}, and essentially confirm their result. As shown in Fig.\ \ref{dcf_go}, we find a peak at zero time-lag, indicating that there is a strong correlation ($\rm DCF_{peak} = 0.95$) with simultaneous flux variations. However, the centroid of the distribution, which provides a higher time resolution, is $\tau_{\rm cen}=-2.5$ days, which means that the $\gamma$-ray variations would follow the optical ones by 2.5 days. We determine the uncertainty in this result by performing Monte Carlo simulations according to the ``flux redistribution/random subset selection" method \citep[FR/RSS;][]{pet98,rai03}, which tests the importance of sampling and data errors. Among the 1000 simulations, about 74\% ($\ga 1 \, \sigma$) of them yield centroids between $-3$ and $-2$ days, so we can conservatively conclude that the delay of the $\gamma$ flux variations after the optical ones is $2.5 \pm 0.5$ days.
If we then run the DCF for the following period, including the 2011 outburst, the result changes completely (see Fig.\ \ref{dcf_go}). The peak is still at $\tau=0$ days, but the correlation weakens ($\rm DCF_{peak} = 0.50$), and the centroid indicates that the optical variations follow the $\gamma$ ones by about 7 days. This result is strongly affected by the presence of the optical flare without $\gamma$-ray counterpart at $\rm JD \sim 2455767$. If we indeed stop the light curves at $\rm JD=2455763$, the resulting DCF still shows a weak correlation, but now the peak and the centroid coincide at $\tau=0$ days.
The reason for the lack of a strong $\gamma$-optical correlation in 2011 has probably to be ascribed to the increase in the optical to $\gamma$-ray flux ratio starting from $\rm JD \sim 2455700$ and culminating with the last ``sterile" optical flare.

We also investigate the possible correlation between the $\gamma$-ray flux and the QSO-corrected optical polarisation $P_0$ derived in Sect.\ \ref{sec_stw}. The corresponding DCF, which is shown in the right panel of Fig.\ \ref{dcf_gogp}, does not provide evidence of any significant correlation.

\section{Pair-production optical depth}

In the emitting plasma jet, $\gamma$-ray photons should be absorbed because of collisions with lower-energy photons, mostly X-rays, which lead to pair production. 
Relativistic beaming of the emitted radiation can explain why we actually observe a $\gamma$-ray flux.
The minimum value of the Doppler factor\footnote{$\delta=[\Gamma_{\rm b} (1-\beta \cos \theta)]^{-1}$, where $\Gamma_{\rm b}$ is the Lorentz factor of bulk motion $\Gamma_{\rm b}=(1-\beta^2)^{-1/2}$, $\beta$ is the ratio of the plasma velocity to the speed of light, and $\theta$ is the viewing angle.} $\delta$ to avoid pair production, i.e.\ to have an optical depth $\tau_{\gamma \gamma}=1$, was derived by \citet{mat93}, and corrected by \citet{mad96}, under the hypothesis that the X-ray and $\gamma$-ray emissions are co-spatial and that the emission region is spherical in the comoving frame
\begin{eqnarray}
\delta_{\rm low}=[ 5 \times 10^{-4} \, (1+z)^{-2 \alpha} \, (1+z-\sqrt{1+z})^{-2} \, h_{75}^2 \, T_5 \times \nonumber \\
(F_{\rm keV}/\mu {\rm Jy})^{-1} \, (E_\gamma/{\rm GeV})^{-\alpha} ]^{-1/(2 \alpha)} \, ,
\end{eqnarray}
where $\alpha$ is the X-ray energy spectral index $F(E)=F_{\rm 1 \, keV}(E/{\rm keV})^{-\alpha}$ whose relationship with the photon spectral index $\Gamma$ is $\alpha=\Gamma-1$, $T_5$ is the flux variability timescale in units of $10^5 \rm \, s$, and $E_\gamma$ is the minimum energy of the observed $\gamma$-ray photons.
By setting $h_{75}=H_0/(75 \rm \, km s^{-1} Mpc^{-1})=0.95$, $\alpha=0.31$, $F_{\rm 1 \, keV}=1.46 \, \rm \mu Jy$ (see Sect.\ \ref{sec_xrt}), $E_\gamma=0.1 \, \rm GeV$, and by considering a variability 
timescale of 12--24 hours, we obtain $\delta_{\rm low}=4.1$--3.7.
This is the lower limit to the Doppler factor affecting the radiation emitted from the jet region where the $\gamma$ and X-ray photons are produced.

\section{Broad-band SED}
\label{sec_sed}

    \begin{figure*}
   \sidecaption
   \includegraphics[width=12cm]{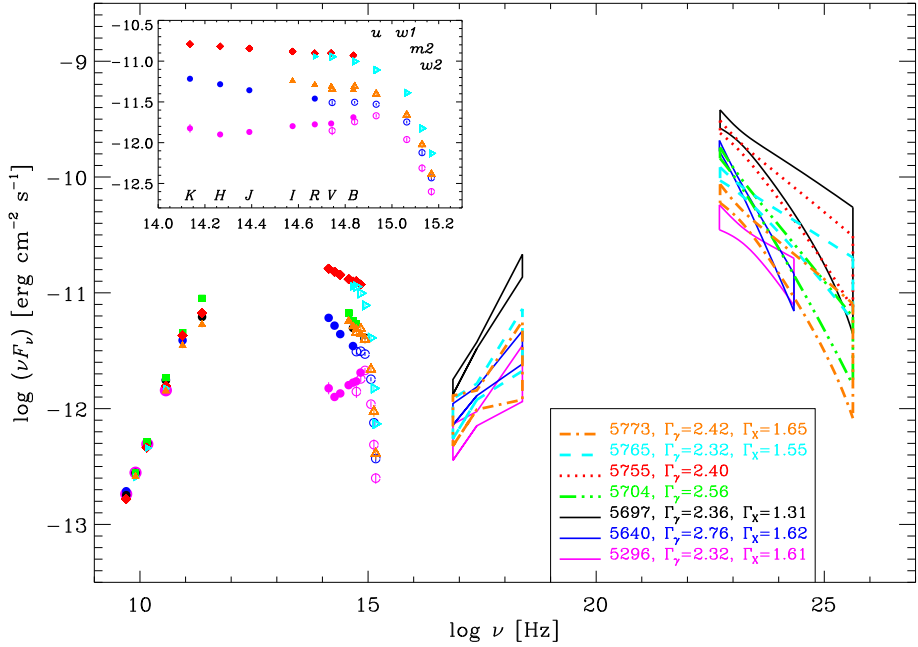}
   \caption{Spectral energy distributions of 4C 38.41 from the radio to the $\gamma$-ray frequencies.
Different colours correspond to different epochs, whose $\rm JD-2450000$ is indicated in the lower inset together with the $\gamma$ and X-ray spectral indices. The near-IR to UV spectral band is enlarged in the upper inset. Filled (empty) circles display ground (space) data. Radio to optical ground data were collected in the framework of the GASP and collaborators; space optical, UV, and X-ray data are from {\it Swift}, $\gamma$-ray data from {\it Fermi}.}
    \label{sed}
    \end{figure*}

Broad-band SEDs of 4C 38.41 are presented in Fig.\ \ref{sed} for seven epochs characterised by different brightness levels and good broad-band data coverage (see also Figs.\ \ref{multi1} and \ref{multi2}).
They are obtained with simultaneous data at near-IR, optical, and higher frequencies, where flux variations are faster, while in the radio band we searched for data within a few days. Near-IR, optical, UV, and X-ray flux densities are corrected for Galactic extinction as explained in the corresponding sections.

In the upper left inset, we show an enlargement of the near-IR to UV frequency range. We note that in the SEDs corresponding to $\rm JD=2455296$ and $\rm JD=2455773$, where ground and space optical data overlap, the space data points are lower than the ground-based data in the $V$ band, especially in the fainter SED.
This agrees with our need in Fig.\ \ref{ottico} to shift the $v$-band UVOT data points by $-0.1$ mag to reconcile them with the ground ones and does not change the conclusion of Sect.\ \ref{sec_uvot} that the thermal-emission contribution likely peaks around the $u$ band.
The shape of the low-energy bump traced by the data shown in Fig.\ \ref{sed} suggests that the synchrotron peak lies in the far-IR--sub-mm frequency range.

The spectral analysis of the X-ray data acquired by {\it Swift} has been presented in Sect.\ \ref{sec_xrt}. 
All the five X-ray spectra shown in Fig.\ \ref{sed} are the result of statistically valid spectral fits, whose spectral indices are indicated in the figure.

To obtain the $\gamma$-ray spectra, we reduced the publicly available {\it Fermi}-LAT data\footnote{{\tt http://Fermi.gsfc.nasa.gov/}} corresponding to the epochs of interest.
The data reduction is based on the unbinned likelihood analysis
with python using the ScienceTools software package version v9r23p1\footnote{\tt {http://fermi.gsfc.nasa.gov/ssc/data/analysis/scitools/\\python\_tutorial.html}}.
We extracted the data from a circular region of interest (RoI) of radius $10\degr$
centred around the location of 4C 38.41 and included events with high
photon probability by keeping only the ``source class'' events (event
class 2). The energy range used in data processing is from 200 MeV to
200 GeV. 
We also filtered the data by excluding photons observed at a zenith
angle of $ > 100\degr$ to reduce contamination from the Earth limb
$\gamma$-rays that result from cosmic rays interacting with the upper
atmosphere.

The $\gamma$-ray flux and spectrum were calculated using the current
public Galactic diffuse emission model (gal\_2yearp7v6\_v0.fits) and the
recommended isotropic model\footnote{\tt {http://fermi.gsfc.nasa.gov/ssc/data/access/lat/\\BackgroundModels.html}}
for the extragalactic diffuse emission (iso\_p7v6source.txt), which
includes the residual cosmic-ray background. The instrument response
function (IRFs) valid for the isotropic spectrum (P7SOURCE\_V6) were
used for data sets with front and back events combined to carry out
the spectral analyses. 

For each of the seven epochs, we calculated
the livetime cube, exposure map, and diffuse response of the
instrument, and applied the algorithm as prescribed by the unbinned
likelihood analysis with python. We used the user contributed tool
(make2FGLxml.py) to correctly model the background by including all of
the sources of interest within the RoI of 4C
38.41, and generated the corresponding XML model file. We carried out
the first spectral fit by setting the spectral model corresponding to
4C 38.41 to ``PowerLaw2'' in the file and keeping the normalisation or
integral factors and photon indices of all sources along with the
normalisation factors of the Galactic diffuse and isotropic components
as free parameters. 
The XML model also includes sources between 10 and
15 degrees from the target, which can contribute to the total counts observed in the RoI owing to the energy-dependent size of the point spread function of the instrument. For these additional sources, normalizations and index were kept fixed to the values of the Second {\it Fermi} LAT Catalog \citep{nol12}.
The first fit was then used to obtain the final
spectral fit after fixing all the free parameters of this fit except
for those of 4C 38.41.

We explored the significance of the $\gamma$-ray signal from the
source using the test statistic (TS) based on the likelihood ratio
test. To ensure a high TS value, we carried out binning using three different
time intervals: (1) 14 days centred around $\rm JD=2455296$, 2455640, 2455704, 2455765,
and 2455773; (2) 3 days centred around $\rm JD=2455697$, and (3) 5 days centred around
$\rm JD=2455755$. The TS value for all seven time intervals are well in excess of
100 ($ > 10 \, \sigma$).

We used the user contributed tool (likeSED.py) to obtain the spectral
points and butterfly plots for each of the seven $\gamma$-ray states.
We created nine customised energy bins to ensure more photons per bin. The tool creates
an exposure map and an event file for each of the energy bins.
We used the final fit from our likelihood analysis to fit
individual bins and, as before, kept parameters of all sources fixed
except for the integral factor and photon index of 4C 38.41. The
source exhibits a slight spectral variation during the seven time
intervals. The individual photon-index values for all seven epochs are displayed in Fig.\ \ref{sed},
with errors indicated by their respective bow-tie patterns.

As displayed in Fig.\ \ref{sed}, the results of the above analysis reveal that the $\gamma$-ray spectra are always very steep and that all epochs are characterised by a strong Compton dominance, as usually found for quasar-like blazars.
This also seems to be true when the jet emission is faint and the thermal contribution prevails in the optical band, as shown by the $\rm JD = 2455296$ SED.
Moreover, by comparing the $\gamma$-ray spectra with the X-ray ones, we can see that higher $\gamma$-ray states correspond to higher X-ray ones. Finally, the data suggest that the peak of the high-energy bump is located in the $10^{21}$--$10^{22}$ Hz frequency interval, i.e.\ in the MeV range.

\section{Discussion}
As discussed in Sect.\ \ref{sec_multi}, in 2009 we found a strong $\gamma$-optical correlation, with a delay in the $\gamma$-ray flux variations after the optical ones of 2--3 days, in agreement with \citet{jor11}. In contrast, in the 2011 outburst the correlation was weak, and the flux changes in the optical band apparently follow those at $\gamma$-rays by about one week.
A similar time lag of the optical flux variations after the $\gamma$-ray ones was found 
by \citet{hay12} when analysing the light curves of another FSRQ, 3C 279. These authors explain the time delay in terms of a decreasing ratio of the external radiation to the magnetic energy densities along the jet, which shifts the location where the optical luminosity reaches a maximum 
downstream from where the $\gamma$ luminosity peaks. As we saw in Sect.\ \ref{sec_sed}, 4C 38.41 also has a strong Compton dominance, which is at the basis of the \citet{hay12} argument, so the same explanation might hold for this source too.

However, our analysis suggests that the apparent optical delay may be due to an increase in the optical flux relative to that in $\gamma$ rays, even leading to an optical flare without a significant $\gamma$-ray counterpart. A similar situation was found by \citet{rai11} for the 2008--2009 outbursts of 3C 454.3.

We obtained a lower limit to the Doppler factor $\delta_{\rm low}=3.7$--4.1 to avoid self-absorption of the $\gamma$ radiation. 
\citet{sav10} combined apparent jet speeds derived from high-resolution VLBA images from the MOJAVE project with flux densities at millimetre wavelengths derived from monitoring data acquired at the Mets\"ahovi Radio Observatory to estimate the jet Doppler factors, Lorentz factors $\Gamma_{\rm b}$, and viewing angles for a sample of 62 blazars. For 4C 38.41 they derived $\beta_{\rm app}=29.5$, $\delta=21.3$, $\Gamma_{\rm b}=31.1$, and $\theta=2.6 \degr$. Since according to \citet{jor11} the $\gamma$-ray emitting region is close to the millimetre one, we can investigate whether the \citet{sav10} findings agree with ours.

If the emission is isotropic in the jet rest-frame and the intrinsic spectrum follows a power-law with index $\alpha$, then the observed flux density $F_\nu (\nu) =\delta^{n+\alpha} \, F'_{\nu'} (\nu)$, where $n=3$ for a discrete, essentially point-like emission region, and $n=2$ for a smooth, continuous jet \citep[e.g.][]{urr95}. Taking into account the uncertainty in both the emitting region structure and the value of $\alpha$, we can assume a flux density dependence on $\delta^3$. 
If the flux density variations that we observe are only due to variations in $\delta$, then 
$\delta  =  \delta_{\rm max}(F_{\nu}/F_{\nu,\rm max})^{1/3}$.

Under the hypotheses that $\delta_{\rm max}=21.3$ from \citet{sav10} and that all the variability shown by the $\gamma$-ray and optical lightcurves in Fig.\ \ref{multi} {\it on a one-week timescale} is due to variations in the Doppler factor, we find that $\delta$ should range between this maximum value and $\delta_{\rm min}=8.0$ to explain the weekly $\gamma$-ray lightcurve.
In the optical band, an uncertainty comes from the subtraction of the QSO contribution derived in Sect.\ \ref{sec_stw} from the $R$-band flux densities. If we consider the weekly-binned $R$-band lightcurve,
$\delta_{\rm min}=6.7$ if $R_{\rm QSO}=17.85$, while $\delta_{\rm min}=7.5$ if $R_{\rm QSO}=18.0$. 
Hence, both the $\gamma$-ray and optical $\delta_{\rm min}$ are consistent with, i.e. greater than, the lower limit $\delta_{\rm low}$ that we obtained.

Moreover, we can also estimate the corresponding variation in the viewing angle $\theta=\rm \arccos \{ [ \Gamma_{\rm b} \, \delta -1]/[ \sqrt{\Gamma_{\rm b}^2-1} \, \delta] \}$ under the assumption that all the variability is of a geometrical nature. Adopting  $\Gamma_{\rm b}=31.1$ after \citet{sav10}, we obtain $\theta_{\rm min}=2.6\degr$ corresponding to $\delta_{\rm max}=21.3$ for both the $\gamma$-ray and optical emitting regions, while we derive $\theta_{\rm max}=4.8$\degr\ for the $\gamma$ radiation, and $\theta_{\rm max}=5.0$ or 5.3\degr\ for the optical when $R_{\rm QSO}=18.0$ or 17.85, respectively. 

In conclusion, the hypothesis that all the flux variability on a weekly timescale is due to changes in the Doppler factor does not contradict the observations. Had we considered shorter variability timescales, we would have obtained lower $\delta_{\rm min}$ values, down to $\delta_{\rm min}=4.3$ in the extreme case that we attributed {\it all} the $R$-band flux density variations to changes in $\delta$ and set $R_{\rm QSO}=17.85$. Even this case is still compatible with the estimated $\delta_{\rm low}$. However, it is likely that the most rapid flux changes are due to intrinsic processes of an energetic nature, such as shocks propagating downstream in the jet. In contrast, had we considered variability timescales longer than a week, we would have obtained weaker constrains. 
Hence, we suggest that Doppler factor variations of a geometric nature provide the most likely explanation of the {\it long-term} flux variability of this source, as already proposed for other blazars \citep[see e.g.][and references therein]{vil09a,lar10,rai11}.

\begin{figure}
   \centering
   \resizebox{\hsize}{!}{\includegraphics{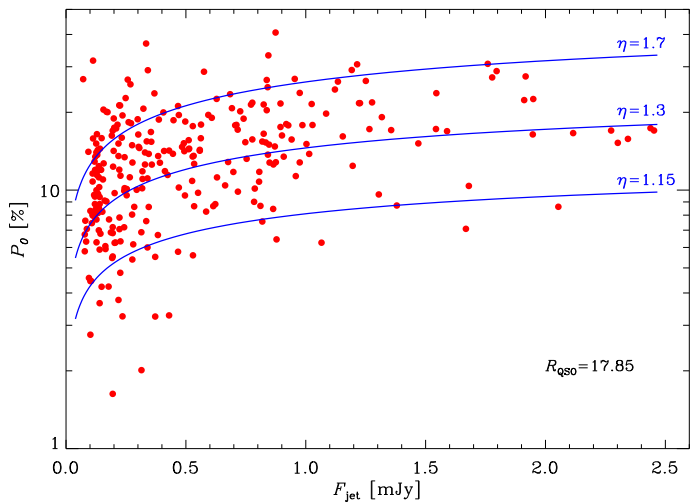}}
   \caption{The intrinsic polarisation of the jet emission $P_0$ derived in Sect.\ \ref{sec_stw} versus the jet flux density in the $R$ band. We have assumed an observed brightness $R_{\rm QSO}=17.85$ to correct for the QSO-like unpolarised emission contribution that affects both the flux and polarisation degree observed. The blue lines represent the results of a shock-in-jet model where the direction of the front wave, hence the polarisation degree, vary because of the variation in the viewing angle. We adopted an optical spectral index of 1.6, as found in Sect.\ \ref{sec_stw}. Values of the shock wave degree of compression $\eta$ in the 1.15--1.7 range can account for most of the data dispersion.}
    \label{hughes}
    \end{figure}

Furthermore, in Sect.\ \ref{sec_stw} we saw that there is a clear dependence of polarisation on brightness (Fig.\ \ref{polar}) that cannot be completely explained in terms of a simple dilution effect from an unpolarised QSO-like emission component (Fig.\ \ref{pola_cor}). 
The intrinsic polarisation variability can be interpreted by various available models.
Here we investigate whether the changes in the viewing angle discussed above to explain the flux variations can also provide a plausible explanation of the variations in the polarisation degree.
If we suppose that for a statistically significant part of time we see shock waves propagating downstream in the jet, because of relativistic aberration the angle between the line of sight and the direction of the normal to the shock-wave front can be expressed as
$$\psi=\arctan({{\sin\theta} \over {\Gamma_{\rm b}\, (\cos\theta-\sqrt{1-\Gamma_{\rm b}^{-2}})}})$$
and the degree of polarisation  $$P_0\approx \frac{\alpha+1}{\alpha+5/3}~\frac{(1-\eta^{-2})\sin^2\psi}{2-(1-\eta^{-2})\sin^2\psi}$$ \citep[e.g.][]{hug85}. 
Here $\eta$ is the degree of compression of the shock wave and $\alpha$ is the optical spectral index (see Sects.\ \ref{sec_opt} and \ref{sec_stw}).
The results of the model are shown in Fig.\ \ref{hughes} for an optical spectral index $\alpha=1.6$ (but they are weakly dependent on $\alpha$) and different choices of $\eta$. They can closely reproduce the behaviour of the intrinsic degree of polarisation of the jet emission $P_0$ derived in Sect.\ \ref{sec_stw} as a function of brightness (here, the intrinsic jet flux density in the $R$ band, which is obtained after correcting the observed flux for the QSO-like contribution and Galactic extinction). Most of the data point dispersion can be accounted for by varying the compression parameter from $\eta=1.15$ to 1.7.
We note that choosing $R_{\rm QSO}=18$, which was obtained as a possible alternative value for the unpolarised emission component in Sect.\ \ref{sec_stw}, would shift the points toward the bottom-right, worsening the agreement between data and model.

\section{Conclusions}
We have presented the results of a huge observing effort on the FSRQ 4C 38.41, carried out by the GASP-WEBT and collaborators from 2007.5 to 2011.9 and by the Steward Observatory from 2008.8 to 2012.1. Earlier data were also collected, so that the optical and radio light curves cover in total about 17 years.
Moreover, we also analysed the UV and X-ray data acquired in 2007--2011 by the {\it Swift} satellite and the $\gamma$-ray data taken in 2008--2011 by {\it Fermi}.
Light curves from the near-IR to the UV band, spanning a factor $\sim 11$ in frequency, show a quite impressive correlation, presumably because this whole range is in the upper part of the synchrotron bump.
In the near-IR-to-UV spectral range, the presence of a QSO-like emission contribution in addition to the synchrotron emission from the jet, is revealed by several findings: 
\begin{itemize}
\item the maximum flux-variation amplitude decreases with increasing frequency;
\item the optical colour shows a redder-when-brighter trend;
\item the optical spectrum reveals unpolarised broad emission-lines with constant flux;
\item the optical polarisation is higher in the red than in the blue;
\item the SEDs display a bump peaking around the $u$ band in faint states.
\end{itemize}
This unpolarised emission component is likely thermal emission from the accretion disc.
Thermal AGN signatures have apparently been found in several blazars based on different information (see e.g. \citealt{per08} for a review, and \citealt{ghi09}), but this is maybe the first time so much observational evidence has been collected for a single object.
Furthermore, we have been able to estimate the brightness of the unpolarised component, $R_{\rm QSO} \sim 17.85$--18, and to correct the observed degree of polarisation for its dilution effect to obtain the polarisation of the jet emission. This still shows a dependence on brightness, which is thus an intrinsic dependence.

The analysis of the radio light curves confirms a scenario where radiation of increasing wavelength is emitted in progressively larger and more external jet regions.
Optical and radio light curves do not show a general correlation, at least on month--year timescales. One possible explanation, which has already been proposed for other blazars, is that the optical and radio emissions come from different jet zones that have variable orientations with respect to the line of sight. Optical and radio flares would thus result when the corresponding emitting regions are more closely aligned, with a consequent enhancement of the Doppler factor.

We found a general correlation between the flux variations in the optical band and at $\gamma$-ray energies, as the optical and $\gamma$-ray periods of increased activity coincide. However, the shape of the $\gamma$-ray outbursts differs from that of the optical ones, and the flux ratios as well as the time lags between variations at the two frequencies change with time. The correlation was strong during 2009, but became weaker in 2011, likely because of an increased activity in the optical band.

We analysed broad-band SEDs of 4C 38.41 built with contemporaneous data in different brightness states. 
A careful spectral analysis of both the X-ray data from {\it Swift} and $\gamma$-ray data from {\it Fermi} was also performed to obtain a reliable spectral shape in the high-energy part of the SED. All the selected epochs show a strong Compton dominance, even when the jet emission is weak and the unpolarised component clearly emerges in the optical band.

We finally discussed a geometrical interpretation of the flux and polarisation variability, which is able to fairly account for the observational data. In this view, at least the long-term flux variations can be ascribed to changes in the Doppler boosting factor produced by changes in the viewing angle. In particular, the weekly $\gamma$-ray and optical light curves would imply changes in $\delta$ and $\theta$ in the ranges of $\sim 7$--21  and $\sim 2.6\degr$--5\degr, respectively. When using these values in the framework of a shock-in-jet model, where the direction of the wave front, and hence the polarisation degree, changes according to the viewing angle, we could also account for the trend of polarisation with brightness.

\begin{acknowledgements}
We thank the referee, Robert Hartman, for his useful comments and suggestions.
We are grateful to Stefano Vercellone for information about the AGILE detection of 4C 38.41.
We acknowledge financial contribution from the agreement ASI-INAF I/009/10/0.
Partly based on observations with the Medicina and Noto telescopes operated by INAF - Istituto di Radioastronomia. 
The work at U. Michigan  was supported by NSF grant AST-0607523 and NASA/Fermi GI grants  NNX09AU16G, NNX10AP16G and NNX11AO13G.
The Submillimeter Array is a joint project between the Smithsonian Astrophysical Observatory and the Academia Sinica Institute of Astronomy and Astrophysics and is funded by the Smithsonian Institution and the Academia Sinica.
This paper is partly based on observations carried out at the German-Spanish Calar Alto Observatory, which is jointly operated by the MPIA and the IAA-CSIC. 
This paper is partly based on observations carried out at the IRAM--30m Telescope, which is supported by INSU/CNRS (France), MPG (Germany), and IGN (Spain). 
Acquisition of the MAPCAT, POLAMI, and MAPI data is partly supported by CEIC (Andaluc\'ia) grant P09-FQM-4784 and by MINECO (Spain) grant and AYA2010-14844.
This work is partly supported by the Georgian National Science foundation grant GNSF/ST09/4-521.
This article is partly based on observations made with the telescopes IAC80 and TCS operated by the Instituto de Astrofisica de Canarias in the Spanish Observatorio del Teide on the island of Tenerife. Most of the observations were taken under the rutinary observation programme. The IAC team acknowledges the support from the  group  of support astronomers and telescope operators of the Observatorio del Teide.
The Steward Observatory spectropolarimetric monitoring project is supported by the {\it Fermi} Guest Investigator grants NNX08AW56G and NNX09AU10G.
Data at SPM observatory were obtained through the support given by PAPIIT grant IN116211.
The Mets\"ahovi team acknowledges the support from the Academy of Finland to our observing projects (numbers 212656, 210338, 121148, and others).
The BU group acknowledges support  by NASA grants  NNX08AV61G, NNX10AU15G, and NNX11AQ03G. 
The PRISM camera at Lowell Observatory was developed by K.\ Janes et al. at BU and Lowell Observatory, with funding from
the NSF, BU, and Lowell Observatory. The Liverpool Telescope is operated on the island of La Palma by Liverpool John Moores University in the Spanish Observatorio del Roque de los Muchachos of the Instituto de Astrofisica de Canarias, with funding from the UK Science and Technology Facilities Council.
The St.\ Petersburg University team acknowledges support from Russian RFBR foundation via grant 12-02-00452.
AZT-24 observations are made within an agreement between  Pulkovo, Rome, and Teramo observatories.
Data at NAO Rozhen were obtained through the support given by the BSF grant DO02 340/08.
We acknowledge the use of public data from the {\it Swift} data archive.
This research has made of the XRT Data Analysis Software (XRTDAS) developed under the responsibility of the ASI Science Data Center (ASDC), Italy.
This research has made use of the SAO/NASA's Astrophysics Data System (ADS) and of
the NASA/IPAC Extragalactic Database (NED), which is operated by the Jet Propulsion Laboratory, California Institute of Technology, under contract with the National Aeronautics and Space Administration. This research has made use of data obtained through the High Energy Astrophysics Science Archive Research Center Online Service, provided by the NASA/Goddard Space Flight Center.
\end{acknowledgements}

\end{document}